\documentclass[sigconf,natbib=true,final]{acmart}

\AtBeginDocument{%
  }

\setcopyright{acmlicensed}
\copyrightyear{2018}
\acmYear{2018}
\acmDOI{XXXXXXX.XXXXXXX}
\acmConference[SIGIR '25]{Make sure to enter the correct
  conference title from your rights confirmation emai}{July 13--18,
2025}{Padova, IT}
\acmISBN{978-1-4503-XXXX-X/18/06}

\usepackage{xspace}
\usepackage{booktabs}
\usepackage{multirow}
\usepackage{pifont}
\usepackage{makecell}
\usepackage{xspace}
\usepackage{pifont}  
\newcommand{\xmark}{\ding{55}} 
\newcommand{\blackcircle}[1]{%
    \raisebox{0.8pt}{\textcircled{\raisebox{-0.8pt}{#1}}}
}
\definecolor{codegreen}{RGB}{93, 168, 128} 
\definecolor{codeblue}{RGB}{74, 74, 250}  
\definecolor{codered}{RGB}{209, 106, 114}
\definecolor{text}{RGB}{68,114,157}
\usepackage{listings}
\usepackage{xcolor}

\definecolor{codepurple}{rgb}{0.58,0,0.82} 
\definecolor{codeorange}{rgb}{0.85,0.33,0.1} 
\definecolor{codeteal}{rgb}{0,0.5,0.5} 
\definecolor{codebackground}{rgb}{0.95,0.98,1} 

\newcommand{\framework}{\textsc{Rankify}\xspace}
\begin{document}

\title[Rankify: A Comprehensive Python Toolkit for Retrieval, Re-Ranking, and RAG]{Rankify: A Comprehensive Python Toolkit  for Retrieval, Re-Ranking, and Retrieval-Augmented Generation}

\author{Abdelrahman Abdallah}
\orcid{0000-0001-8747-4927}
\affiliation{%
  \institution{University of Innsbruck}
  \city{Innsbruck}
  \state{Tyrol}
  \country{Austria}
  }
\email{abdelrahman.abdallah@uibk.ac.at}

\author{Bhawna Piryani}
\orcid{0009-0005-3578-2393}
\affiliation{%
  \institution{University of Innsbruck}
  \city{Innsbruck}
  \state{Tyrol}
  \country{Austria}
  }
\email{bhawna.piryani@uibk.ac.at}

\author{Jamshid Mozafari}
\orcid{0000-0003-4850-9239}
\affiliation{%
  \institution{University of Innsbruck}
 \city{Innsbruck}
  \state{Tyrol}
  \country{Austria}
  }
\email{jamshid.mozafari@uibk.ac.at}

\author{Mohammed Ali}
\affiliation{%
  \institution{University of Innsbruck}
  \city{Innsbruck}
  \state{Tyrol}
  \country{Austria}
  }
\email{Mohammed.ali@uibk.ac.at	}

\author{Adam Jatowt}
\orcid{0000-0001-7235-0665}
\affiliation{%
  \institution{University of Innsbruck}
  \city{Innsbruck}
 \state{Tyrol}
  \country{Austria}
  }
\email{adam.jatowt@uibk.ac.at}

\renewcommand{\shortauthors}{Abdallah et al.}


\begin{abstract}
Retrieval, re-ranking, and retrieval-augmented generation (RAG) are critical components of modern applications in information retrieval, question answering, or knowledge-based text generation. However, existing solutions are often fragmented, lacking a unified framework that easily integrates these essential processes. The absence of a standardized implementation, coupled with the complexity of retrieval and re-ranking workflows, makes it challenging for researchers to compare and evaluate different approaches in a consistent environment. While existing toolkits such as Rerankers and RankLLM provide general-purpose reranking pipelines, they often lack the flexibility required for fine-grained experimentation and benchmarking.  In response to these challenges, we introduce \textbf{\framework}, a powerful and modular open-source toolkit designed to unify retrieval, re-ranking, and RAG within a cohesive framework. \framework supports a wide range of retrieval techniques, including dense and sparse retrievers, while incorporating state-of-the-art re-ranking models to enhance retrieval quality. Additionally, \framework includes a collection of pre-retrieved datasets to facilitate benchmarking, available at Huggingface\footnote{\url{https://huggingface.co/datasets/abdoelsayed/reranking-datasets-light}}. To encourage adoption and ease of integration, we provide comprehensive documentation\footnote{\url{http://rankify.readthedocs.io/}}, an open-source implementation on GitHub\footnote{\url{https://github.com/DataScienceUIBK/rankify}}, and a PyPI package for easy installation\footnote{\url{https://pypi.org/project/rankify/}}. As a unified and lightweight framework, \framework allows researchers and practitioners to advance retrieval and re-ranking methodologies while ensuring consistency, scalability, and ease of use.
\end{abstract}



\received{20 February 2007}
\received[revised]{12 March 2009}
\received[accepted]{5 June 2009}

\maketitle

\begin{figure}[!t]
  \centering
  \includegraphics[width=0.25\columnwidth]{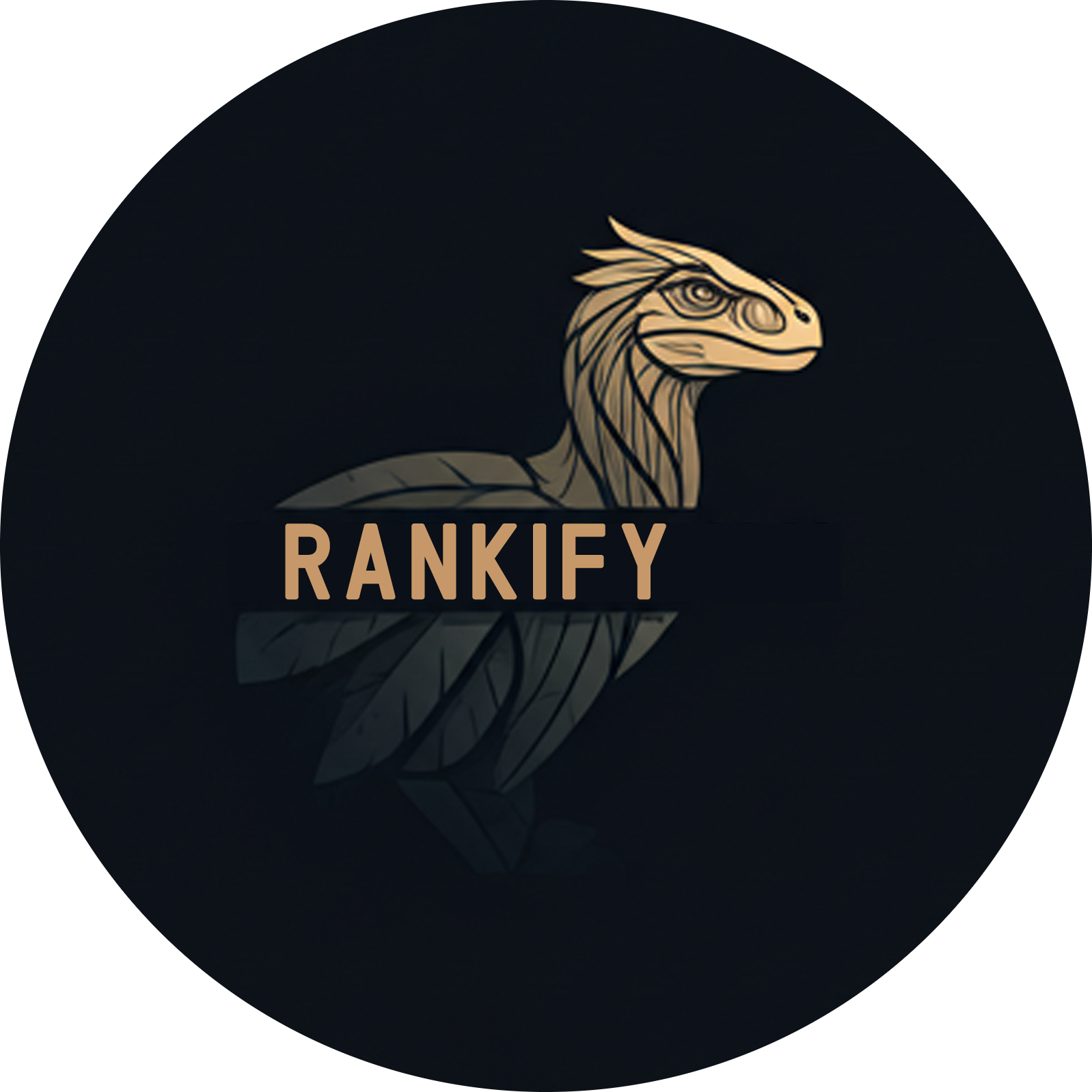}
  \caption{\framework logo.}
  \label{fig:rankify_logo}
\end{figure}

\section{Introduction}

Information retrieval (IR) systems~\cite{singhal2001modern,chowdhury2010introduction} are fundamental to many  applications~\cite{wang2024utilizing,liu2024information}, including question-answering~\cite{khamnuansin2024mrrank,abdallah-etal-2025-dynrank}, search engines~\cite{croft2010search,xiong2024search}, and knowledge-based generation~\cite{long2024generative}. These systems often rely on a two-stage pipeline: a retriever that efficiently identifies a set of candidate documents and a re-ranker that refines these results to maximize relevance to the query~\cite{dpr,monobert,rocketqav2}. This approach has proven highly effective, with retrieval and re-ranking methods achieving state-of-the-art performance across diverse NLP benchmarks. 

\begin{table*}[ht]
\caption{Comparison of Retriever, Re-ranking, and RAG toolkits. Modular Design indicates if the toolkit uses modular components. Automatic Evaluation refers to the availability of built-in evaluation tools. Corpus shows whether the toolkit provides utilities for corpus processing, such as cleaning and chunking. The Combination column represents the total number of possible configurations, calculated as the product of the number of datasets, retrievers, re-rankers, and RAG methods (e.g., for \framework: 40 × 7 × 24 × 3 = 20,160).}
\centering
\resizebox{1.0\textwidth}{!}{
\begin{tabular}{lcccccccc}
\toprule
\textbf{Toolkit} & \textbf{\# Datasets} & \textbf{\# Retrievers} & \textbf{\# Re-rankers} & \textbf{\# RAG Methods} & \textbf{Modular Design} & \textbf{Automatic Evaluation} & \textbf{Corpus} & \textbf{Combination}\\
\midrule
FlashRAG~\cite{jin2024flashrag} & 32 & 3 & 2 & 12 & \checkmark & \checkmark & \checkmark&2,304 \\
FastRAG~\cite{abane2024fastrag} & 0 & 1 & 3 & 7 & \checkmark & \xmark & \xmark&  21\\
AutoRAG~\cite{kim2024autoragautomatedframeworkoptimization} & 4 & 2 & 7 & 1 & \checkmark & \checkmark & \xmark&56 \\
LocalRQA~\cite{yu2024localrqa} & 0 & 2 & 0 & 0 & \xmark & \checkmark & \xmark & 2\\
Rerankers~\cite{clavi2024rerankers} & 0 & 0 & 15 & 0 & \checkmark & \checkmark & \xmark&15 \\
CHERCHE~\cite{sourty2022cherche} & 0 & 7 & 2 & 0 & \xmark & \xmark & \xmark & 14\\
\textbf{\framework} & \textbf{40} & \textbf{7} & \textbf{24} & \textbf{3} & \checkmark & \checkmark & \checkmark &20,160\\
\bottomrule
\end{tabular}
}

\label{tab:framework_comparison}
\end{table*}
Retrievers form the backbone of information retrieval systems by identifying a subset of documents relevant to a user's query. The retrieval landscape includes sparse, dense, and hybrid methods. Sparse retrievers, such as BM25~\cite{bm25}, rely on exact term matching by representing queries and documents as high-dimensional sparse vectors. These methods are highly effective when query terms closely align with document content but they struggle to capture semantic relationships. Dense retrievers, including models like DPR~\cite{dpr}, BPR~\cite{yamada2021bpr}, ColBERT~\cite{santhanam2021colbertv2}, BGE~\cite{bge_m3}, Contriever~\cite{izacard2021contriever} and ANCE~\cite{ance}, overcome this limitation by encoding queries and documents into low-dimensional dense vectors using neural networks, enabling retrieval based on semantic similarity even when lexical overlap is minimal. However, dense retrieval requires significant computational resources for training and inference. Hybrid retrieval methods~\cite{yang2019hybrid,chen2022out} integrate the strengths of both approaches, combining the precision of sparse retrievers with the semantic understanding of dense models to achieve a more balanced and robust retrieval system.

Re-ranking techniques enhance the initial retrieval results by ensuring the most relevant documents appear at the top. These methods are categorized into three main approaches:
(1) Pointwise reranking~\cite{monobert,abdallah-etal-2025-dynrank,abdallah2025asrankzeroshotrerankinganswer,sachan2022improving} treats reranking as a regression or classification task, assigning independent relevance scores to each document.
(2) Pairwise reranking~\cite{huang2024pairdistill,qin2023large,sinhababu2024few} refines ranking by comparing document pairs and optimizing their relative order based on relevance. However, this approach treats all document pairs equally, sometimes improving lower-ranked results at the expense of top-ranked ones.
(3) Listwise reranking~\cite{rankgpt,pradeep2023rankzephyr,yoon2024listt5listwisererankingfusionindecoder} considers the entire document list, prompting LLMs with the query and a subset of candidates for reranking. Due to input length limitations, these methods tend to employ a sliding window strategy, progressively refining rankings from back to front.

Retrieval-Augmented Generation (RAG)~\cite{lewis2020retrieval,gao2023retrieval} enhances language models by integrating retrieval and generation, making them more effective in knowledge-intensive tasks. Rather than relying solely on pre-trained knowledge, RAG dynamically retrieves relevant documents from external sources during inference, incorporating them into the generation process. 

The growing complexity and diversity of retrieval, re-ranking, and RAG methods pose significant challenges for benchmarking, reproducibility, and integration. Existing toolkits, such as Pyserini~\cite{Lin_etal_SIGIR2021_Pyserini}, Rerankers~\cite{clavi2024rerankers} and RankLLM~\cite{pradeep2023rankvicunazeroshotlistwisedocument} often lack flexibility, enforce rigid implementations, and require extensive preprocessing, making them less suitable for research-driven experimentation. Additionally, retrieval and re-ranking datasets are scattered across different sources, complicating evaluation and comparison. To address these challenges, we introduce \framework, an open-source framework that unifies retrieval, re-ranking, and RAG into a modular and extensible ecosystem (logo shown in Figure \ref{fig:rankify_logo}). \framework supports diverse retrieval techniques, integrates the state-of-the-art re-ranking models, and provides curated pre-retrieved datasets to streamline experimentation. Designed for maximizing flexibility, it enables researchers to efficiently build, evaluate, and extend retrieval pipelines while ensuring consistency in benchmarking. 

The main contributions of this paper are as follows:

\begin{itemize} 

\item \textbf{Curated retrieval datasets and precomputed embeddings:} \framework provides 40 datasets, each with 1,000 pre-retrieved documents per query, across various domains (QA, dialogue, entity linking, etc.). It also includes pre-computed Wikipedia and MS MARCO corpora for multiple retrievers, eliminating preprocessing overhead.

\item \textbf{Diverse retriever and re-ranking support:} \framework integrates dense (DPR, ANCE, BPR, ColBERT, BGE, Contriever) and sparse (BM25) retrievers, along with 24 state-of-the-art re-ranking models, enabling seamless retrieval and ranking experimentation.

\item \textbf{RAG integration and evaluation tools:} \framework bridges retrieval, re-ranking, and RAG by passing retrieved documents to LLMs for evaluation. 

\item  \textbf{Comprehensive evaluation tools:} \framework offers a diverse range of evaluation metrics and tools for retrieval and question answering. The framework is accompanied by extensive online documentation, making it easy for users to explore its features. Additionally, \framework is freely available on PyPI and GitHub, ensuring accessibility for researchers and practitioners.

\end{itemize}

\begin{figure*}
    \centering
    \includegraphics[width=0.8\linewidth]{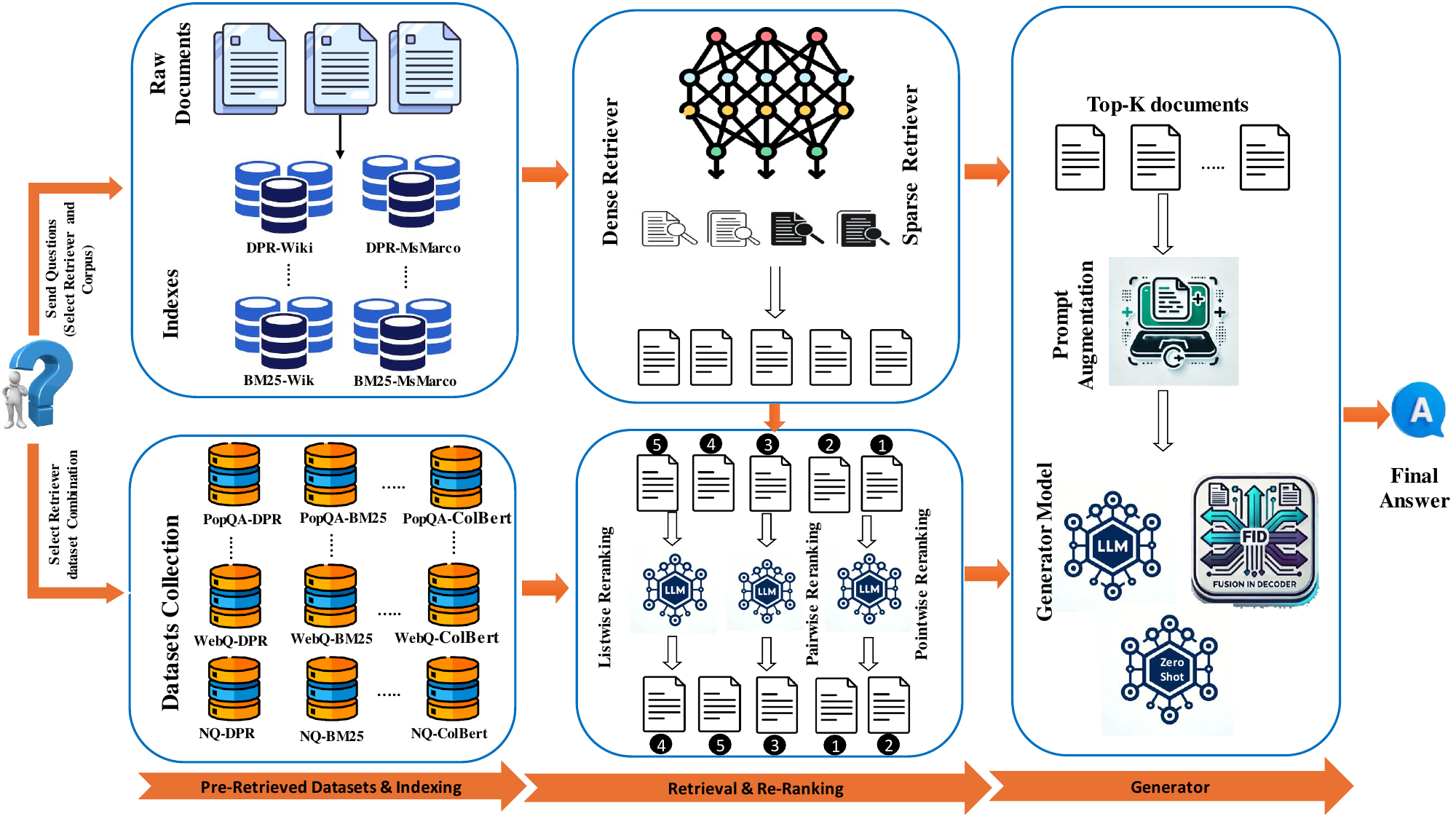}
    \caption{An overview of the Rankify pipeline, demonstrating its dual capability for document retrieval. Users can interact with the system either by providing a query to retrieve relevant documents in real-time or by leveraging pre-retrieved datasets already indexed by the framework. The process starts with Pre-Retrieved Datasets \& Corpus Indexing, where documents are indexed using both dense (e.g., DPR) and sparse (e.g., BM25) retrievers across corpus like Wikipedia and MS MARCO. Next, in the Retrieval \& Re-Ranking stage, the system retrieves candidate documents using dense, or sparse retrieval methods and re-ranks them with pointwise pairwise, or listwise models powered by large language models (LLMs). Finally, the Generator stage applies prompt augmentation and uses models like Fusion-in-Decoder (FiD) to generate accurate and contextually informed answers.}
    \label{fig:workflow}
\end{figure*}

\section{Related Work}\label{s:related_work}
Research in information retrieval (IR), re-ranking, and retrieval-augmented generation (RAG) has progressed significantly over the past decade. Traditional retrieval models like BM25~\cite{bm25} offered robust lexical matching capabilities but struggled with capturing semantic relationships. This limitation led to the development of dense retrieval methods~\cite{lee-etal-2019-latent}, which leverage pre-trained neural encoders to represent queries and documents in a shared semantic space. Notable approaches, such as DPR~\cite{dpr}, ANCE~\cite{ance}, and multi-vector models like ColBERT~\cite{khattab2020colbertefficienteffectivepassage}, have demonstrated substantial improvements in retrieval effectiveness. Hybrid retrievers, combining sparse and dense signals~\cite{DBLP:conf/ecir/GaoDCFDC21,ma2021replication}, further enhance performance by leveraging both lexical and semantic features. Recent advancements, including knowledge distillation~\cite{qu-etal-2021-rocketqa} and curriculum learning~\cite{cldrd}, continue to refine retrieval performance across diverse datasets.

Re-ranking methods have also evolved alongside retrieval techniques to improve the ordering of retrieved documents. Traditional pointwise~\cite{zhu2021leveraging} and pairwise~\cite{cao2007learning} approaches have given way to listwise methods like LambdaRank and ListNet~\cite{burges2010ranknet,liu2017listnet}. Deep neural models, including cross-encoders and transformer-based architectures, have demonstrated remarkable success in re-ranking tasks by capturing complex interactions between queries and documents. Zero-shot and in-context re-ranking with large language models (LLMs), such as GPT-4~\cite{achiam2023gpt} and RankT5~\cite{zhuang2023rankt5}, now enable effective ranking adjustments without task-specific training.

Retrieval-Augmented Generation (RAG) has emerged as a powerful paradigm for enhancing generative models in knowledge-intensive tasks~\cite{lewis2020retrieval}. By retrieving relevant documents and integrating them into the generative process, RAG systems improve factual accuracy and reduce hallucinations. Techniques like self-consistency~\cite{wang2022self} and noise filtering~\cite{fang2024enhancing} have been proposed to further improve the reliability of RAG outputs. However, the effectiveness of these systems heavily depends on the quality of retrieved and re-ranked documents, highlighting the need for robust retrieval frameworks.

In response to the growing complexity of IR, re-ranking, and RAG tasks, several frameworks have been introduced. \textit{Rerankers}~\cite{clavi2024rerankers} provides a lightweight Python interface for common re-ranking models, while \textit{RankLLM} focuses on listwise re-ranking with LLMs. Other frameworks like \textit{FlashRAG}~\cite{jin2024flashrag} and \textit{AutoRAG}~\cite{kim2024autoragautomatedframeworkoptimization} offer modular components for RAG experimentation, though they often lack support for diverse datasets and advanced retriever configurations. Tools such as LangChain~\cite{chase2022langchain}, LlamaIndex~\cite{Liu_LlamaIndex_2022}, and DSPy~\cite{khattab2023dspy} further contribute to the ecosystem by simplifying model integration and workflow design.  A comparison of retrieval, re-ranking, and RAG toolkits is presented in Table~\ref{tab:framework_comparison}, highlighting the advantages of \framework in dataset diversity, retriever and re-ranker support, and modularity.

\begin{figure}[t]
  \centering
  \includegraphics[width=0.90\columnwidth]{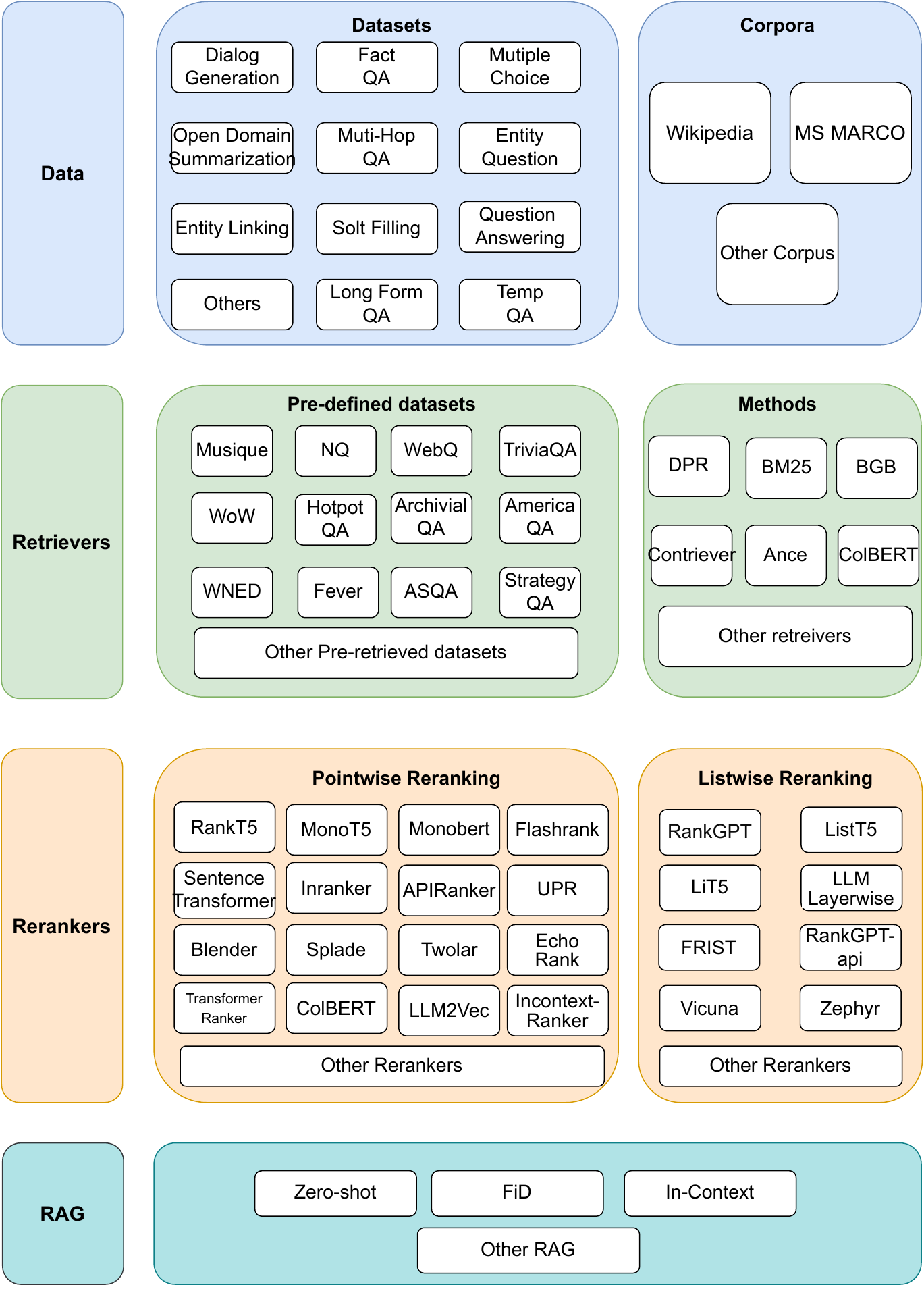}
  \caption{The architecture of the \framework, showing the interplay between its core modules: \textit{Datasets}, \textit{Retrievers}, \textit{Re-Rankers}, and \textit{RAG Evaluation}. Each module operates independently while seamlessly integrating with others, enabling end-to-end retrieval and ranking workflows.}
  \label{fig:rankify_framework}
\end{figure}

\section{\framework}\label{s:rankify}

Modern information retrieval systems rely on a combination of retrieval, ranking, and generative techniques to surface relevant content and generate accurate responses. However, existing solutions are often fragmented, requiring researchers and practitioners to integrate multiple toolkits to experiment with different retrieval and ranking strategies. \framework addresses this challenge by offering an end-to-end retrieval and ranking ecosystem, enabling seamless experimentation across a wide range of retrieval, re-ranking, and retrieval-augmented generation (RAG) models.

Unlike existing retrieval and ranking frameworks that focus on a single aspect of the pipeline, \framework is designed to be modular, extensible, and lightweight, allowing users to plug in different retrievers, ranking models, and RAG techniques with minimal effort. Built in Python, it leverages machine learning libraries such as TensorFlow~\cite{10.5555/3026877.3026899}, PyTorch~\cite{10.5555/3454287.3455008}, Spacy~\cite{honnibal2020spacy},  and Pyserini~\cite{Lin_etal_SIGIR2021_Pyserini}  . The framework is available on PyPI\footnote{\url{https://pypi.org/project/rankify/}}, making installation straightforward:

\begin{lstlisting}[numbers=none]
$ pip install rankify
\end{lstlisting}

\framework supports both precomputed and real-time retrieval workflows, allowing users to benchmark with pre-retrieved datasets or perform live document retrieval on large-scale corpora. The framework includes ready-to-use indices for Wikipedia and MS MARCO, eliminating the need for time-consuming indexing.  

The architecture of \framework consists of four core modules:  
\begin{itemize}
    \item \textbf{Datasets}: Provides standardized access to datasets like Natural Questions, TriviaQA, and HotpotQA, with support for custom dataset creation.  
    \item \textbf{Retrievers}: Integrates diverse retrieval methods such as BM25, DPR, ANCE, BGE, and ColBERT, enabling flexible retrieval strategies.  
    \item \textbf{Re-Rankers}: Includes 24 models with 41 sub-methods, supporting pointwise, pairwise, and listwise re-ranking.  
    \item \textbf{RAG}: Facilitates retrieval-augmented generation with models like LLAMA, GPT, and T5, supporting in-context learning, Fusion-in-Decoder (FiD), and Incontext RALM.  
\end{itemize}

A key advantage of \framework is its unified API, which abstracts implementation details and simplifies experimentation across different models. Figure~\ref{fig:workflow} and ~\ref{fig:rankify_framework} illustrates the interactions between these modules within the pipeline. The framework is scalable, adaptable, and suited for both research and real-world applications like search engines and question-answering systems.  \framework is open-source, actively maintained\footnote{\url{https://github.com/abdoelsayed2016/reranking}}, and accompanied by comprehensive documentation\footnote{\url{http://rankify.readthedocs.io/}} to support easy adoption for both beginners and experts.

\begin{table}
\caption{Summary of datasets used in \framework.  Categories include QA, Multi-Hop QA, Long-Form QA, Multiple-Choice, Entity-Linking, Slot Filling, Fact Verification, Dialog Generation, Summarization, and Other specialized datasets.}
\small
\centering

\resizebox{0.45\textwidth}{!}{
\begin{tabular}{l l c c c}  
\toprule
    Task & Dataset Name & \# Train & \# Val & \# Test \\ 
\midrule
    \multirow{14}{*}{{\textbf{QA} }} & NQ~\cite{naturalquestion}& 79,168 & 8,757 & 3,610 \\ 
    &TriviaQA~\cite{triviaqa} &  78,785 & 8,837 & 11,313 \\ 
    &WebQ~\cite{webquestions}  & 3,778 & - & 2,032 \\ 
    &SQuAD~\cite{squad} &  87,599 & 10,570 & - \\ 
    &NarrativeQA~\cite{narrativeqa} &  32,747 & 3,461 & 10,557 \\ 

    &MSMARCO-QA~\cite{msmarco} &  808,731 & 101,093 & - \\ 
     &PopQA~\cite{popqa} & - & - & 14,267 \\ 

    &SIQA~\cite{siqa} & 33,410 & 1,954 & - \\ 
    &Fermi~\cite{fermi} &  8,000 & 1,000 & 1,000 \\ 
    &WikiQA~\cite{wikiqa}  & 20,360 & 2,733 & 6,165 \\ 
    &AmbigQA~\cite{ambigqa,naturalquestion} &  10,036 & 2,002 & - \\ 

    &CommenseQA~\cite{commonsenseqa}  & 9,741 & 1,221 & - \\ 
    &PIQA~\cite{piqa} & 16,113 & 1,838 & - \\ 
    &BoolQ~\cite{boolq} &  9,427 & 3,270 & - \\

\midrule
    \multirow{4}{*}{{\textbf{Multi-Hop QA} }} &2WikiMultiHopQA~\cite{2wikimultihop}  & 15,000 & 12,576 & - \\ 
    &Bamboogle~\cite{selfask_2023}  & - & - & 125 \\ 
    &Musique~\cite{musique}  & 19,938 & 2,417 & - \\ 

     &HotpotQA~\cite{hotpotqa} &  90,447 & 7,405 & - \\ 

\midrule
   \multirow{2}{*}{{\textbf{Temp QA} }} & ArchivalQA~\cite{wang2022archivalqa} & 384,426 & 48,304 & 48760 \\ 
    &ChroniclingQA~\cite{piryani2024chroniclingamericaqa}& 385,629   & 21,739    & 21,735   \\
\midrule
   \multirow{2}{*}{{\textbf{Long-Form QA} }}  &ELI5~\cite{eli5} &  272,634 & 1,507 & - \\
   
   & ASQA~\cite{asqa} & 4,353 & 948 & - \\ 
    
\midrule
    \multirow{5}{*}{{\textbf{Multiple-Choice} }}& MMLU~\cite{mmlu,mmlu_ethics}  & 99,842 & 1,531 & 14,042 \\ 
    &TruthfulQA~\cite{truthfulqa} & - & 817 & - \\ 
    &HellaSwag~\cite{hellaswag}  & 39,905 & 10,042 & - \\ 
    &ARC~\cite{arc_challenge}  & 3,370 & 869 & 3,548 \\ 
    &OpenBookQA~\cite{OpenBookQA2018}  & 4,957 & 500 & 500 \\ 
\midrule
   \multirow{2}{*}{{\textbf{Entity-linking} }}  
   &WNED~\cite{wned,kilt_2021}  & - & 8,995 & - \\ 
   & AIDA CoNLL-YAGO~\cite{AIDA_CONLL,kilt_2021}  & 18,395 & 4,784 & - \\ 
    
\midrule
    \multirow{2}{*}{{\textbf{Slot filling} }} 
    & Zero-shot RE~\cite{levy-etal-2017-zero,kilt_2021}  & 147,909 & 3,724 & - \\ 

    &T-REx~\cite{trex,kilt_2021}  & 2,284,168 & 5,000 & - \\ 

\midrule
    \multirow{-1}{*}{{\textbf{Dialog Generation} }} & WOW~\cite{dinan2018wizard,kilt_2021} & 63,734 & 3,054 & - \\ 

\midrule
    \multirow{1}{*}{{\textbf{Fact Verification} }} &FEVER~\cite{fever,kilt_2021} & 104,966 & 10,444 & - \\ 
\midrule
    \multirow{1}{*}{{\textbf{\text{Open-domain Summarization}} }} &WikiAsp~\cite{wikiasp}  & 300,636 & 37,046 & 37,368 \\ 
\bottomrule
\end{tabular}
}
\label{tab:Dataset-size}
\end{table}

\subsection{Prebuilt Retrieval Corpora and Indexes}

\framework provides two large-scale retrieval corpora: \textbf{Wikipedia}~\cite{karpukhin2020dense} and \textbf{MS MARCO}~\cite{msmarco}, enabling researchers to conduct retrieval experiments on well-established benchmarks. To support efficient retrieval, \framework includes pre-built index files for each retriever, allowing users to directly query the corpora without requiring costly indexing. Each corpus has been indexed for multiple retrieval models, including BM25, DPR, ANCE, BGE, Contriever, and ColBERT. By providing precomputed indexes, \framework simplifies large-scale retrieval research while maintaining consistency across different retrieval pipelines. Users can view all available datasets using the following command (see Listing~\ref{lst:available_datasets}):
\begin{lstlisting}[language=Python, caption={This script retrieves the latest information about the available datasets, and displays metadata for each dataset in the terminal.}, label={lst:available_datasets} , captionpos=b]
from rankify.dataset.dataset import Dataset
Dataset.avaiable_dataset()
\end{lstlisting}

\subsection{Datasets}\label{s:dataset}
\framework provides a standardized and efficient way to handle datasets for retrieval, re-ranking, and retrieval-augmented generation (RAG). To simplify retrieval workflows, \framework integrates pre-retrieved datasets with structured annotations, enabling users to experiment with various retrieval and re-ranking methods with minimal code. The datasets in \framework are categorized based on task type and size, as shown in Table~\ref{tab:Dataset-size}. Each dataset follows a standardized schema with three main components: the \textbf{Question}, which represents the query requiring relevant information; the \textbf{Answers}, which are the expected correct responses to the query; and the \textbf{Contexts} (or \textbf{Retrieved Documents)}, which provide a ranked list of candidate documents retrieved by a specific retriever.

To support benchmarking and reproducibility, \framework provides pre-retrieved datasets, each with 1,000 top-ranked documents from Wikipedia\footnote{Note: We are currently processing different retrievers based on  MS MARCO to generate and store 1,000 top-ranked documents per query for each dataset.}, processed by various retrievers such as BM25, DPR, ANCE, BGE, Contriever, and ColBERT. These datasets cover diverse Re-Ranking and RAG  tasks.

\framework allows users to load and explore datasets with minimal effort. Below is an example demonstrating how to \textbf{download and inspect a dataset} using \framework’s API:

\begin{lstlisting}[language=Python, caption={Loading a dataset in Rankify and inspecting its structure.}, captionpos=b]
from rankify.dataset.dataset import Dataset
# Load the nq-test dataset retrieved with BM25
dataset = Dataset(retriever="bm25", dataset_name="nq-test")
documents = dataset.download(force_download=False)
\end{lstlisting}


Beyond preloaded datasets, \framework enables users to define their own datasets using the Dataset class. Below is an example demonstrating how to create a custom retrieval dataset:

\begin{lstlisting}[language=Python, caption={Creating a custom dataset in Rankify.}, label={lst:custom_dataset}, captionpos=b]
from rankify.dataset.dataset import Dataset, Document, Question, Answer, Context
question = Question("What is the capital of France?")
answer = Answer(["Paris"])
context = [Context(score=0.9, has_answer=True, id=1, title="France", text="The capital is Paris.")]
document = Document(question, answer, context)
custom_dataset = Dataset(retriever="custom", dataset_name="MyDataset")
custom_dataset.documents = [document]
custom_dataset.save_dataset("my_dataset.json")
\end{lstlisting}

Users can load their own datasets using \texttt{load\_dataset} and \texttt{load\_dataset\_qa} functions. The \texttt{load\_dataset} function is designed for structured datasets that include both queries and retrieved documents. On the other hand, \texttt{load\_dataset\_qa} is intended for question-answering datasets that contain only queries and answers, allowing users to use them directly for retrieval tasks. The following examples demonstrate how users can load these datasets in \framework:

\begin{lstlisting}[language=Python, caption={Loading a dataset with documents and a QA-only dataset.}, label={lst:load_dataset}, captionpos=b]
from rankify.dataset.dataset import Dataset
retrieval_dataset = Dataset.load_dataset("path/to/retrieval_dataset.json")
qa_dataset = Dataset.load_dataset_qa("path/to/qa_dataset.jsonl")
\end{lstlisting}

\subsection{Retriever Models}\label{ss:retrievers}

\framework supports a diverse set of retriever models to facilitate document retrieval for various information retrieval tasks. 
The current version includes both sparse retrievers, such as BM25~\cite{bm25}, and dense retrievers, such as DPR~\cite{dpr}, ANCE~\cite{ance}, and BGE~\cite{bge_m3}, Contriever~\cite{izacard2021contriever} and ColBERT~\cite{khattab2020colbertefficienteffectivepassage}. The retrievers implemented in \framework are accessible via a unified interface, where users can specify the retrieval method and parameters. 

\framework allows users to retrieve documents efficiently using different retrieval models. The retrievers can be applied to any dataset supported by \framework, making it flexible for experimentation and benchmarking. Users can easily initialize a retriever and apply it to a list of query-document pairs, as demonstrated in Listing~\ref{lst:retriever_usage}.

\begin{lstlisting}[language=Python, caption={Using a retriever model in Rankify to retrieve documents.}, label={lst:retriever_usage}, captionpos=b]
from rankify.dataset.dataset import Document, Question, Answer
from rankify.retrievers.retriever import Retriever

# Define sample documents
documents = [
    Document(question=Question("the cast of a good day to die hard?"), 
             answers=Answer(["Jai Courtney", "Bruce Willis"]), contexts=[]),
    Document(question=Question("Who wrote Hamlet?"), 
             answers=Answer(["Shakespeare"]), contexts=[])
]
# Initialize a retriever (example: ColBERT)
retriever = Retriever(method="colbert", model="colbert-ir/colbertv2.0", n_docs=5, index_type="msmarco")
# Retrieve documents
retrieved_documents = retriever.retrieve(documents)
# Print retrieved documents
for i, doc in enumerate(retrieved_documents):
    print(f"\nDocument {i+1}:")
    print(doc)
\end{lstlisting}

In Listing~\ref{lst:retriever_usage}, users can specify the retrieval corpus by setting \texttt{index\_type} to \texttt{msmarco} or \texttt{wiki}, select the desired retriever model, and define the number of documents to retrieve. This flexibility allows researchers to experiment with different retrieval settings, optimizing their retrieval strategy based on the task requirements.

\subsection{Re-Ranking Models}\label{ss:rerankers}

In information retrieval, two-stage pipelines are widely used to maximize retrieval performance. The first stage involves retrieving a set of candidate documents using a computationally efficient retriever, followed by a second stage where these documents are re-ranked using a stronger, typically, neural network-based model. This re-ranking step enhances retrieval quality by considering deeper semantic relationships between the query and documents. While effective, re-ranking models vary significantly in their architectures, trade-offs, and implementations, making it challenging for users to select the most suitable method for their specific needs.

To address this, \framework provides a unified interface for re-ranking, enabling users to effortlessly switch between different models with minimal modifications. The framework supports a diverse set of 24 primary re-ranking models with 41 sub-methods, spanning different re-ranking strategies such as pointwise, pairwise, and listwise approaches. The implemented models include MonoBERT~\cite{nogueira2019passage}, MonoT5~\cite{nogueira2020document}, RankT5~\cite{zhuang2023rankt5}, ListT5~\cite{yoon2024listt5listwisererankingfusionindecoder}, ColBERT~\cite{santhanam2021colbertv2}, RankGPT~\cite{rankgpt}, and various transformer-based re-rankers.

Users can apply these models to re-rank retrieved documents using a simple interface, as demonstrated in Listing~\ref{lst:reranker_usage}.

\begin{lstlisting}[language=Python, caption={Applying re-ranking in Rankify.}, label={lst:reranker_usage}, captionpos=b]
from rankify.dataset.dataset import Document, Question, Answer
from rankify.retrievers.retriever import Retriever
from rankify.rerankers.reranker import Reranker
# Define sample documents
documents = [
    Document(question=Question("the cast of a good day to die hard?"), 
             answers=Answer(["Jai Courtney", "Bruce Willis"]), contexts=[]),
    Document(question=Question("Who wrote Hamlet?"), 
             answers=Answer(["Shakespeare"]), contexts=[])
]
# Initialize a retriever
retriever = Retriever(method="colbert", model="colbert-ir/colbertv2.0", n_docs=5, index_type="msmarco")
retrieved_documents = retriever.retrieve(documents)
# Initialize a re-ranker
reranker = Reranker(method="monot5",model_name="monot5-base")
# Apply re-ranking
reranked_documents = reranker.rerank(retrieved_documents)
# Print re-ranked documents
for i, doc in enumerate(reranked_documents):
    print(f"\nDocument {i+1}:")
    print(doc)
\end{lstlisting}

\framework supports both pre-trained models from Hugging Face and API-based re-ranking models such as RankGPT, Cohere, and Jina Reranker. Users can select from a variety of models, including cross-encoders, T5-based ranking models, and late-interaction retrieval models like ColBERT. The flexibility to switch between different re-ranking strategies allows researchers and practitioners to optimize ranking performance based on their task requirements.

\subsection{Retrieval-Augmented Generation (RAG) Models}\label{ss:rag}

Retrieval-Augmented Generation (RAG) enhances language models by integrating retrieval mechanisms, allowing them to generate responses based on dynamically retrieved documents rather than relying solely on pre-trained knowledge. This approach is particularly effective for knowledge-intensive tasks such as open-domain question answering, fact verification, and knowledge-based text generation. \framework provides a modular and extensible interface for applying multiple RAG methods, including zero-shot generation, Fusion-in-Decoder (FiD)~\cite{izacard2020leveraging}, and in-context learning ~\cite{ram-etal-2023-context}.

In \framework, the Generator module enables seamless integration of RAG techniques, allowing users to experiment with different generative approaches. Users can specify the desired RAG method and model, applying generation strategies across retrieved documents. 

Users can apply these methods to generate responses based on retrieved documents. Listing~\ref{lst:generator_usage} demonstrates how to use \framework's RAG module with an in-context learning approach.

\begin{lstlisting}[language=Python, caption={Applying Retrieval-Augmented Generation (RAG) in Rankify.}, label={lst:generator_usage}, captionpos=b]
from rankify.dataset.dataset import Document, Question, Answer, Context
from rankify.generator.generator import Generator
# Sample question and contexts
question = Question("What is the capital of France?")
answers = Answer(["Paris"])
contexts = [
    Context(id=1, title="France", text="The capital of France is Paris.", score=0.9),
    Context(id=2, title="Germany", text="Berlin is the capital of Germany.", score=0.5)
]
# Create a Document
doc = Document(question=question, answers=answers, contexts=contexts)
# Initialize Generator with In-Context RALM
generator = Generator(method="in-context-ralm", model_name='meta-llama/Llama-3.1-8B')
# Generate answer
generated_answers = generator.generate([doc, doc])
print(generated_answers)
\end{lstlisting}

\framework allows users to leverage large-scale language models such as LLaMA~\cite{touvron2023llama}, GPT-4~\cite{achiam2023gpt}, and T5-based models~\cite{raffel2020exploring} for retrieval-augmented generation. By supporting both encoder-decoder architectures (FiD~\cite{izacard2020leveraging}) and decoder-only models (e.g., GPT, LLaMA), the framework provides flexibility for optimizing generation quality based on task-specific requirements.

\subsection{Evaluation Metrics}\label{ss:evaluation}

\begin{table*}[h!]
\caption{Retrieval performance of BM25 across several selected datasets. The table reports Top-1, Top-5, Top-10, Top-20, Top-50, and Top-100 retrieval accuracy for each dataset split. Datasets vary in complexity, including open-domain QA (e.g., TriviaQA, NQ), multi-hop reasoning (e.g., 2WikiMultiHopQA, HotpotQA), fact verification (e.g., TruthfulQA), and temporal retrieval (e.g., ArchivalQA, ChroniclingAmericaQA).}
\centering
\resizebox{0.90\textwidth}{!}{
\begin{tabular}{@{}lccccccc|lccccccc@{}}
\toprule
Dataset & Split & Top-1 (\%) & Top-5 (\%) & Top-10 (\%) & Top-20 (\%) & Top-50 (\%) & Top-100 (\%) & Dataset & Split & Top-1 (\%) & Top-5 (\%) & Top-10 (\%) & Top-20 (\%) & Top-50 (\%) & Top-100 (\%) \\ \midrule
\multirow{2}{*}{2WikiMultiHopQA~\cite{2wikimultihop}}
& Train & 8.85 & 19.78 & 28.15 & 38.16 & 51.16 & 57.70
& \multirow{2}{*}{ArchivalQA~\cite{wang2022archivalqa}}
& Val & 18.18 & 33.29 & 39.72 & 45.96 & 53.68 & 58.74 \\ 
& Val & 17.17 & 31.72 & 40.14 & 47.73 & 55.88 & 61.17
&
& Test & 17.68 & 32.55 & 39.26 & 45.36 & 52.76 & 58.08 \\ 
\midrule

\multirow{2}{*}{ChroniclingAmericaQA~\cite{piryani2024chroniclingamericaqa}}
& Val & 4.26 & 11.80 & 16.42 & 21.50 & 28.19 & 33.15
& \multirow{2}{*}{AmbigQA~\cite{ambigqa,naturalquestion}}
& Train & 27.30 & 50.67 & 59.63 & 67.63 & 75.68 & 79.87 \\ 
& Test & 4.18 & 11.23 & 15.83 & 20.87 & 27.49 & 32.48
&
& Val & 30.22 & 54.25 & 65.58 & 72.98 & 80.47 & 84.92 \\ 
\midrule

\multirow{3}{*}{ARC~\cite{arc_challenge}}
& Train & 4.45 & 18.16 & 31.39 & 51.39 & 78.19 & 91.36
& \multirow{3}{*}{TriviaQA~\cite{triviaqa}}
& Train & 47.87 & 67.22 & 72.90 & 77.15 & 81.30 & 83.64 \\ 
& Val & 4.83 & 19.45 & 30.84 & 52.13 & 79.06 & 91.83
&
& Val & 48.70 & 67.57 & 72.84 & 77.38 & 81.19 & 83.59 \\ 
& Test & 4.96 & 19.84 & 33.57 & 51.94 & 79.09 & 91.80
&
& Test & 48.22 & 67.44 & 72.79 & 77.28 & 81.42 & 83.87 \\ 
\midrule

\multirow{3}{*}{SQuAD~\cite{squad}}
& Train & 31.84 & 50.24 & 57.15 & 63.33 & 70.32 & 74.61
& \multirow{3}{*}{MMLU~\cite{mmlu,mmlu_ethics}}
& Train & 3.65 & 14.14 & 23.98 & 38.68 & 62.22 & 78.04 \\ 
& Val & 36.94 & 57.61 & 64.78 & 71.31 & 77.89 & 81.96
&
& Val & 6.53 & 21.75 & 33.64 & 50.62 & 75.77 & 88.18 \\ 
& Test & 36.67 & 57.37 & 64.50 & 71.09 & 77.74 & 81.84
&
& Test & 6.53 & 22.01 & 34.67 & 51.22 & 74.35 & 86.40 \\ 
\midrule

\multirow{3}{*}{QuaRTz~\cite{tafjord-etal-2019-quartz}}
& Train & 7.81 & 26.82 & 40.62 & 63.54 & 84.38 & 92.19
& \multirow{3}{*}{NarrativeQA~\cite{narrativeqa}}
& Train & 16.80 & 25.24 & 28.46 & 31.50 & 35.74 & 38.65 \\ 
& Val & 7.81 & 26.82 & 40.62 & 63.54 & 84.38 & 92.19
&
& Val & 15.46 & 23.92 & 27.10 & 30.45 & 35.08 & 38.89 \\ 
& Test & 5.87 & 23.60 & 38.90 & 58.67 & 80.74 & 90.43
&
& Test & 16.80 & 25.24 & 28.46 & 31.50 & 35.74 & 38.65 \\ 
\midrule

\multirow{3}{*}{NQ~\cite{naturalquestion}}
& Train & 22.89 & 44.97 & 54.32 & 62.73 & 71.75 & 77.03
& \multirow{3}{*}{OpenBookQA~\cite{OpenBookQA2018}}
& Train & 3.35 & 16.24 & 28.99 & 47.85 & 75.01 & 88.56 \\ 
& Val & 23.62 & 45.60 & 55.41 & 64.19 & 72.51 & 77.72
&
& Val & 3.40 & 15.20 & 31.60 & 50.60 & 79.00 & 90.00 \\ 
& Test & 23.46 & 45.62 & 56.32 & 64.88 & 74.57 & 79.72
&
& Test & 4.00 & 16.80 & 32.20 & 51.20 & 74.00 & 88.80 \\ 
\midrule

\multirow{2}{*}{WebQuestions~\cite{webquestions}}
& Train & 21.41 & 44.71 & 55.35 & 64.21 & 72.71 & 78.45
& \multirow{2}{*}{Musique~\cite{musique}}
& Train & 6.23 & 13.70 & 18.54 & 24.39 & 33.77 & 40.81 \\ 
& Test & 19.54 & 42.86 & 53.44 & 63.25 & 72.34 & 76.43
&
& Val & 7.36 & 15.47 & 21.39 & 25.98 & 34.59 & 41.29 \\ 
\midrule

\multirow{1}{*}{T-REx~\cite{trex,kilt_2021}}
& Val & 45.48 & 64.34 & 70.60 & 76.04 & 80.42 & 82.98
& \multirow{1}{*}{TruthfulQA~\cite{truthfulqa}}
& Val & 2.33 & 7.59 & 12.12 & 20.69 & 38.19 & 56.55 \\ 
\midrule

\multirow{2}{*}{SIQA~\cite{siqa}}
& Train & 0.34 & 1.52 & 2.78 & 4.85 & 8.92 & 13.43
& \multirow{2}{*}{HotpotQA~\cite{hotpotqa}}
& Train & 34.89 & 51.71 & 58.13 & 64.09 & 71.11 & 75.49 \\ 
& Val & 0.61 & 2.05 & 3.58 & 6.55 & 11.31 & 16.58
&
& Val & 28.13 & 45.02 & 51.92 & 57.79 & 65.44 & 70.67 \\ 
\midrule

\multirow{1}{*}{StrategyQA~\cite{geva-etal-2021-aristotle}}
& Train & 4.19 & 15.72 & 26.29 & 39.87 & 52.88 & 57.86
& \multirow{1}{*}{PopQA~\cite{popqa}}
& Test & 25.00 & 38.77 & 44.70 & 49.74 & 56.82 & 62.36 \\ 
\midrule

\multirow{2}{*}{WOW~\cite{dinan2018wizard,kilt_2021}}
& Train & 0.26 & 0.38 & 0.41 & 0.45 & 0.49 & 0.52
& \multirow{2}{*}{ZSRC~\cite{levy-etal-2017-zero}}
& Train & 50.53 & 67.23 & 72.06 & 76.10 & 80.35 & 82.99 \\ 
& Val & 0.20 & 0.33 & 0.36 & 0.39 & 0.46 & 0.46
&
& Val & 52.26 & 70.86 & 76.29 & 81.44 & 85.39 & 87.73 \\ 
\midrule

\multirow{1}{*}{Bamboogle~\cite{selfask_2023}}
& Test & 5.60 & 12.80 & 17.60 & 24.80 & 39.20 & 44.00
& \multirow{1}{*}{EntityQuestions~\cite{sciavolino2021simple}}
& Test & 43.36 & 60.50 & 66.08 & 70.61 & 75.41 & 79.06 \\

\bottomrule
\end{tabular}
}
\label{tab:bm25_results}
\end{table*}

\framework provides evaluation metrics for retrieval, re-ranking, and retrieval-augmented generation (RAG). For retrieval and re-ranking, Top-k accuracy measures whether documents containing the correct answer appear within the top-k results. This definition of relevance, while task-specific, aligns with the goal of downstream applications like question answering. Listing~\ref{lst:evaluation_retrieval} demonstrates how to compute Top-k accuracy.

\begin{lstlisting}[language=Python, caption={Computing Top-k accuracy for retrieval and re-ranking in Rankify.}, label={lst:evaluation_retrieval}, captionpos=b]
from rankify.metrics.metrics import Metrics
metrics = Metrics(documents)
# Compute Top-k accuracy before and after re-ranking
before_ranking = metrics.calculate_retrieval_metrics(ks=[1,5,10,20,50,100], use_reordered=False)
after_ranking = metrics.calculate_retrieval_metrics(ks=[1,5,10,20,50,100], use_reordered=True)
print("Before Ranking:", before_ranking)
print("After Ranking:", after_ranking)
\end{lstlisting}

For QA and RAG, \framework supports Exact Match (EM), recall, precision, and containment to evaluate generated answers. Listing~\ref{lst:evaluation_rag} shows how these metrics are computed.

\begin{lstlisting}[language=Python, caption={Computing QA and RAG evaluation metrics in Rankify.}, label={lst:evaluation_rag}, captionpos=b]
from rankify.metrics.metrics import Metrics
qa_metrics = Metrics(documents)
qa_results = qa_metrics.calculate_generation_metrics(generated_answers)
print(qa_results)
\end{lstlisting}

By providing a unified evaluation module, \framework ensures consistent benchmarking across retrieval, re-ranking, and RAG tasks.

\section{Experimental Result and Discussion}

\subsection{Experiment Setup}

\framework enables researchers to benchmark retrieval, re-ranking, and RAG methods, evaluate their own approaches, and explore optimizations within these tasks. To demonstrate its capabilities, we conducted multiple experiments to provide reproducible benchmarks and performance insights. All main experiments were conducted on 2x NVIDIA A100 GPUs.

Experiments were performed on a diverse set of datasets, covering open-domain and multi-hop QA, fact verification, and temporal retrieval tasks. All the experiments in our study use preprocessed
English Wikipedia dump from December 2018 as
released by~\cite{dpr} as evidence passages. Each Wikipedia article is split into non-overlapping 100-word passages, with over 21 million passages in total.

We evaluated retrieval performance on datasets such as Natural Questions (NQ), TriviaQA, HotpotQA, 2WikiMultiHopQA, and ArchivalQA, while re-ranking performance was analyzed on MSMARCO, WebQuestions, and PopQA. For retrieval evaluation, we measured Top-k accuracy at k={1, 5, 10, 20, 50, 100}. Exact match (EM), Precision, Recall, Contains, and F1 were used as the primary evaluation metrics for QA tasks.

We conducted experiments across all supported retrieval, re-ranking, and RAG methods. Retrieval models BM25, DPR, ANCE, BGE, and ColBERT. Re-ranking methods tested included MonoT5, RankT5, RankGPT, and various transformer-based re-rankers. For RAG, we evaluated zero-shot generation, Fusion-in-Decoder (FiD), in-context learning (RALM). These experiments highlight the adaptability of \framework for benchmarking retrieval, ranking, and knowledge-grounded text generation.

\begin{table}[!ht]
\caption{Comparison of different retrieval models on NQ, WebQ, and TriviaQA. The table reports Top-1 to Top-100 accuracy for each retriever.}
\scriptsize
\centering
\begin{tabular}{@{}llcccccc@{}}
\toprule
Retriever & Dataset & Top-1 & Top-5 & Top-10 & Top-20 & Top-50 & Top-100 \\
\midrule




\multirow{3}{*}{Contriever~\cite{izacard2021contriever}} & NQ              & 38.81 & 65.65 & 73.91 & 79.56 & 84.88 & 88.01 \\
           & WebQ          & 35.97 & 63.83 & 69.78 & 75.94 & 81.64 & 83.76 \\
           & TriviaQA        & 49.85 & 71.39 & 76.68 & 80.26 & 83.85 & 85.72 \\

\midrule

\multirow{3}{*}{DPR~\cite{dpr}}        & NQ              & 44.57 & 67.76 & 74.52 & 79.50 & 84.40 & 86.81 \\
           & WebQ            & 44.64 & 63.98 & 70.52 & 75.05 & 80.51 & 82.97 \\
           & TriviaQA        & 57.47 & 72.40 & 76.50 & 79.76 & 82.96 & 85.09 \\
\midrule
\multirow{3}{*}{ColBert~\cite{santhanam2021colbertv2}}        & NQ  & 42.99 & 68.78 & 76.12 & 82.08 & 86.15 & 88.59 \\
           & WebQ    & 40.11 & 64.81 & 71.56 & 76.57 & 81.20 & 84.50 \\
           & TriviaQA            & 57.36 & 75.90 & 79.49 & 82.18 & 84.74 & 86.41 \\
\midrule
\multirow{3}{*}{Ance~\cite{ance}}        & NQ       &       50.80 & 71.86 & 78.12 & 82.52 & 86.23 & 88.28 \\
           & WebQ                & 46.51 & 66.24 & 71.80 & 76.62 & 81.79 & 84.25 \\
           & TriviaQA           & 56.46 & 72.14 & 76.65 & 80.04 & 83.37 & 85.22\\

\midrule
\multirow{3}{*}{BGE~\cite{bge_embedding,bge_m3}}        & NQ              & 48.03 & 72.22 & 78.50 & 82.66 & 86.90 & 89.45  \\
           & WebQ                & 42.67 & 65.45 & 72.54 & 79.04 & 82.78 & 85.88  \\
           & TriviaQA            & 57.81 & 75.39 & 79.83 & 82.92 & 85.53 & 86.94\\

\bottomrule
\end{tabular}

\label{tab:combined_qa}
\end{table}

\begin{table}[h]
\caption{Top-20/Top-100 retrieval accuracy of different retrievers on three open-domain QA datasets (NQ, TriviaQA, and WebQ). Our \framework implementation achieves results identical to Pyserini across all retrievers, as we use Pyserini as the indexing backend. This validates the correctness of our implementation. We also match the official GitHub results for ColBERT and Contriever on these datasets. 
}

\centering\scalebox{.80}{
\begin{tabular}{llll}
\hline
        Retrievers & NQ          & TriviaQA           & WebQ \\
\midrule
DPR~\cite{dpr}      & 78.4 / 85.4 & 79.4 / 85.0  & 73.2 / 81.4 \\
DPR\_Pyserini~\cite{Lin_etal_SIGIR2021_Pyserini} &  79.5 /  86.8 & 79.7 / 85.1  & 75.1 / 82.9 \\
DPR\_\framework &  79.5 /  86.8 & 79.7 / 85.1  & 75.1 / 82.9 \\
\hline
BM25~\cite{sachan2022improving}     &  63.0	 / 78.2 & 76.4	  / 83.1  &   62.3 /  75.5 \\

BM25\_Pyserini~\cite{Lin_etal_SIGIR2021_Pyserini}    &  64.8  / 79.7  & 77.3 / 83.8   &  63.3 / 76.4  \\

BM25\_\framework &  64.8  / 79.7  & 77.3 / 83.8   &  63.3 / 76.4  \\
\hline
Contriever~\cite{contriever}      & 79.6 / 88.0 &  80.4	/ 85.7  &  75.9 /   83.7\\
Contriever\_\framework &   79.6 / 88.0 &  80.3 / 85.7  & 75.9 /  83.7\\
\hline

ColBert~\cite{santhanam2021colbertv2}     &    82.1 / 88.5 &  82.2 / 86.4  &   76.6 / 84.5 \\
ColBert\_\framework  &  82.1 / 88.5 &  82.2 / 86.4   &   76.6 / 84.5  \\
\hline

\hline

ANCE~\cite{ance}    &     82.1 / 87.9 & 80.3 / 85.2  &   - / - \\
ANCE\_\framework   &    82.5 / 88.5& 80.0 / 85.2  &   76.6 / 84.25 \\
\hline
\end{tabular}}

\label{tab_dpr}

\end{table}

\begin{table}[!ht]
\caption{Performance of re-ranking methods on BM25-retrieved documents for NQ Test and WebQ Test. Results are reported in terms of Top-1, Top-5, Top-10, Top-20, and Top-50 accuracy, highlighting the impact of various re-ranking models on retrieval effectiveness. Please note that some results may differ from the original papers (e.g., UPR) as our experiments were conducted with the top 100 retrieved documents, whereas the original studies used 1,000 documents for ranking.
}
\small
\centering
\resizebox{0.40\textwidth}{!}{
\begin{tabular}{@{}l |l| c c c  | c c c@{}}\toprule
Reranking/ &  Model  &  \multicolumn{3}{c}{NQ} & \multicolumn{3}{c}{WebQ}  \\

& & Top-1 & Top-10 & Top-50 & Top-1 & Top-10 & Top-50   \\
\midrule
BM25      & - & 23.46 & 56.32 & 74.57 & 19.54  & 53.44 & 72.34 \\

\midrule
 
\multirow{9}{*}{UPR~\cite{sachan2022improving}}  & T5-small& 23.60 & 59.97  & 75.15 &  18.55 &  55.56   &  72.68  \\

  & T5-base & 26.81 & 63.32 & 76.12 & 20.66 & 58.56 & 72.68  \\
  & T5-large& 29.75 & 65.67 & 76.48 & 24.21 & 60.38 & 73.32 \\
  & T0-3B & 35.42 & 67.56 & 76.75 & 32.48 & 64.17 & 73.67 \\
  & gpt2 & 25.95 & 60.47 & 75.87 & 20.47 & 56.49 & 72.78 \\
  & gpt2-medium &26.75 & 63.04 & 75.95 & 22.39 & 59.54 & 72.44 \\
  &gpt2-large & 26.59 &  62.68  & 75.95 & 24.06   &  59.84  & 72.78  \\

  & gpt2-xl & 27.28  & 63.24 & 75.84 &  23.57   & 60.48 &  72.73  \\
  &gpt-neo-2.7B &28.75  & 64.81  &  76.56 &24.75    & 59.64 &  72.63  \\

\midrule

 \multirow{1}{*}{RankGPT~\cite{rankgpt}} & llamav3.1-8b & 41.55 & 66.17 & 75.42 & 38.77 & 62.69 & 73.12 \\
 
\midrule

 \multirow{4}{*}{FlashRank~\cite{Damodaran2024FlashRank}} 
  & TinyBERT-L-2-v2& 31.49 & 61.57 & 74.95 & 28.54 & 60.62 & 73.17 \\
  & MultiBERT-L-12 & 11.99&  43.54 & 69.63 & 12.54  & 45.91 & 67.91 \\
  & ce-esci-MiniLM-L12-v2 & 34.70 & 64.81 & 76.17 & 31.84 & 62.54 & 73.47  \\
 & T5-flan & 7.95  & 36.14  & 66.67 &   12.05 & 42.96   & 67.27  \\
\midrule
  
  \multirow{3}{*}{RankT5~\cite{zhuang2023rankt5}}
  & base & 43.04 & 68.47 & 76.28 & 36.95 & 64.27 & 74.45  \\
  & large &45.54 & 70.02 & 76.81 & 38.77 & 66.48 & 74.31 \\
  &3b & 47.17& 70.85 & 76.89 & 40.40 & 66.58 & 74.45  \\

\midrule

\multirow{3}{*}{Inranker~\cite{laitz2024inranker}}  
& small & 15.90 & 46.84 & 69.83  & 14.46  & 46.25 & 69.98 \\
&base  &15.90 & 48.11 & 69.66 &14.46 & 46.80 & 69.68 \\
&3b &15.90 & 48.06 & 69.00 & 14.46 & 46.11 & 69.34 \\
\midrule

\multirow{1}{*}{LLM2Vec~\cite{behnamghader2024llm2vec}} 
&Meta-Llama-31-8B & 24.32 & 59.55 & 75.26  &  26.72 &60.48 & 73.47\\

\midrule

\multirow{1}{*}{MonoBert~\cite{monobert}} 
& large & 39.05 & 67.89 & 76.56 & 34.99 &64.56 & 73.96 \\

\midrule

\multirow{1}{*}{Twolar~\cite{baldelli2024twolar}} 
& twolar-xl & 46.84 & 70.22 & 76.86 &41.68 &67.07 &74.40 \\

\midrule

\multirow{2}{*}{Echorank~\cite{rashid2024ecorank}} 
& flan-t5-large& 36.73 & 59.11 & 62.38 & 31.74 &58.75 &61.51 \\
&flan-t5-xl&   41.68 & 59.05 & 62.38  & 36.22 &57.18 &61.51\\

\midrule

\multirow{2}{*}{\makecell[l]{Incontext\\ Reranker~\cite{chen2024attentionlargelanguagemodels}}} &  \multirow{2}{*}{llamav3.1-8b} & \multirow{2}{*}{15.15} & \multirow{2}{*}{57.11}  & \multirow{2}{*}{76.48}  & \multirow{2}{*}{18.89}   & \multirow{2}{*}{52.16} &  \multirow{2}{*}{71.70} \\
&&&&&&&\\

\midrule

\multirow{6}{*}{Lit5~\cite{tamber2023scaling}}
&  LiT5-Distill-base  & 40.05 & 65.95 & 75.73  & 36.76 & 63.48 & 73.12 \\
&  LiT5-Distill-large  & 44.40 & 67.59 & 76.01   &  39.66  &64.56 &73.67\\
&  LiT5-Distill-xl  &  47.81 & 68.55 & 76.26  &  42.37  &65.55 &73.62 \\
& LiT5-Distill-base-v2   &  42.57 &  66.73 & 75.56  & 39.61 &64.22 &73.32\\
&  LiT5-Distill-large-v2  & 46.53 &  67.83 & 75.87  & 41.97 &65.64 &72.98 \\
&  LiT5-Distill-xl-v2  & 47.92 & 69.03 & 76.17  & 41.53 &65.69 &73.27  \\

\midrule

\multirow{10}{*}{ \makecell[l]{Sentence \\ Transformer \\ Reranker} }
& GTR-base~\cite{ni2021large} & 39.41 &65.95 & 76.03  & 36.56 &64.32 &73.62  \\
& GTR-large & 40.63 &68.25 & 76.73  & 38.97 &65.30 &73.57 \\
& T5-base~\cite{raffel2020exploring} & 31.19 &63.60 & 76.06  & 29.77 &62.84 &73.52 \\
& T5-large & 30.80 &63.35 &76.37   & 30.51  &61.71 &73.37 \\
&   all-MiniLM-L6-v2~\cite{wang2020minilmv2} & 33.35  & 65.37  &  76.01 & 30.95   & 62.10  & 73.52 \\
& GTR-xl   & 41.55  & 67.78  & 76.81 &  38.92   &  66.04  & 74.01 \\
& GTR-xxl   & 42.93 &  68.55&  77.00 &39.41    & 65.89 & 74.01 \\
&  T5-xxl  & 38.89  &  67.78  &76.64   &  35.82  & 65.20  & 74.01 \\
&  Bert-co-condensor  & 30.96  & 61.91   &  75.20 &  32.43  &  62.20  & 73.08 \\
&  Roberta-base-v2  & 32.60  & 63.24  & 75.42  &  31.34  &  62.64 &  73.37 \\

\bottomrule           
\end{tabular}}

\label{tab:reranking-result}
\end{table}


\subsection{Retrieval Results}

The retrieval performance of BM25 across multiple datasets is shown in Table~\ref{tab:bm25_results}. The results report Top-1, Top-5, Top-10, Top-20, Top-50, and Top-100 retrieval accuracy for each dataset split. Performance varies significantly based on dataset characteristics, with high Top-k accuracy achieved on open-domain QA datasets such as TriviaQA and Natural Questions. On the other hand, more complex multi-hop datasets like 2WikiMultiHopQA and HotpotQA are characterized by lower retrieval rates, reflecting the challenges in retrieving supporting evidence across multiple documents. Temporal datasets like ArchivalQA and ChroniclingAmericaQA exhibit moderate retrieval performance, likely due to the temporal nature of their content. It is important to note that some datasets in Table~\ref{tab:bm25_results} were either not officially recorded in previous studies or rely on different underlying corpora, making direct comparison difficult. Additionally, some datasets (e.g., ChroniclingAmericaQA) do not report retrieval accuracy as a standard Top-k metric, further complicating direct validation. Despite these variations, our results shown in Table~\ref{tab:bm25_results} provide a comprehensive overview of BM25's strengths and limitations. 

Table~\ref{tab:combined_qa} presents the retrieval performance of various dense retrievers across three benchmark datasets: NQ, WebQ, and TriviaQA. The results report Top-1, Top-5, Top-10, Top-20, Top-50, and Top-100 retrieval accuracy, providing insights into how different retrieval models rank relevant documents. Among the tested retrievers, dense retrieval methods such as DPR, MSS-DPR, and ColBERT consistently outperform retrievers like MSS and Contriever. MSS-DPR achieves the highest Top-1 accuracy, reaching 50.16\% on NQ, 44.24\% on WebQ, and 61.63\% on TriviaQA, demonstrating the effectiveness of fine-tuned dense retrieval. Standard DPR also performs well, particularly on NQ and TriviaQA, where it reaches 48.67\% and 57.47\% Top-1 accuracy, respectively. ColBERT, which employs late-interaction ranking, shows competitive performance, particularly in higher recall settings (Top-50 and Top-100).

\begin{figure*}[h!]
    \centering
    \includegraphics[width=0.75\linewidth]{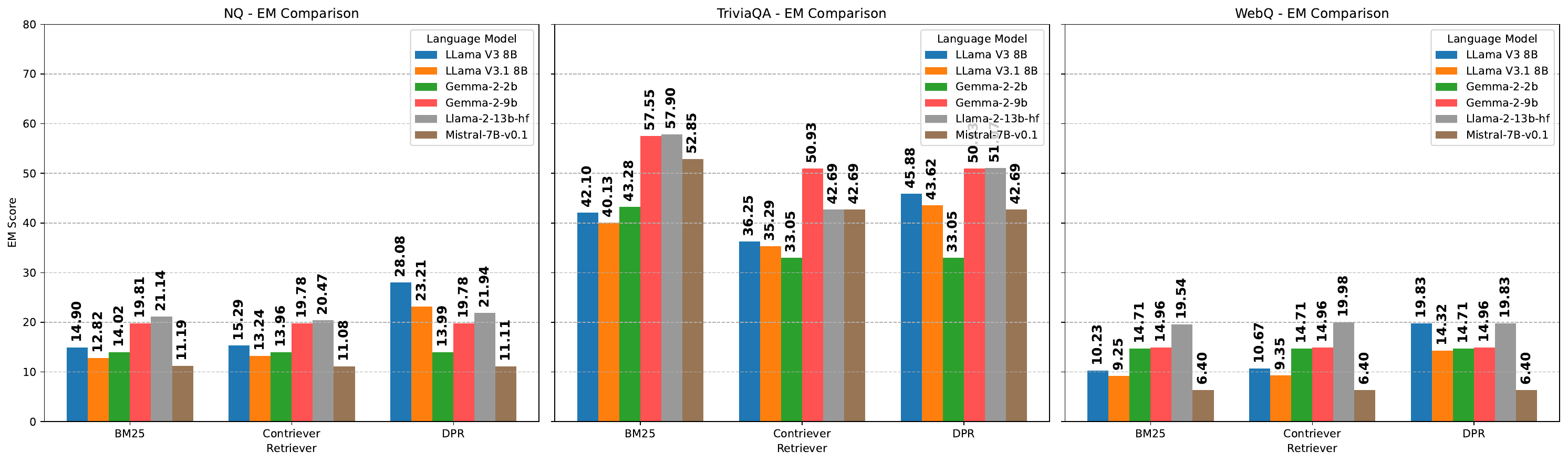}
    \caption{Exact Match (EM) for BM25, Contriever, and DPR retrievers across three datasets (NQ, TriviaQA, WebQ) using various language models (LLaMA V3/V3.1, Gemma 2B/9B, LLaMA 2 13B, and Mistral 7B).}
    \label{fig:generator}
\end{figure*}

\subsection{Comparison with Original Implementations}

To validate the correctness and reliability of the \framework's retrieval module, we compare its performance against original implementations and Pyserini-based baselines for multiple retrievers. Specifically, we evaluate DPR, Contriever, BM25, and ColBERT across three open-domain QA datasets:  NQ, TriviaQA, and WebQ. The retrieval performance is measured using Top-20 and Top-100 accuracy.

Table~\ref{tab_dpr} presents the results of our \framework implementation alongside the original models and Pyserini-based implementations. Our results show that \framework achieves retrieval performance identical to Pyserini for DPR, BM25 as we utilize Pyserini as the backend for indexing and retrieval.  However, for Contriever BGE, and ColBERT, we implemented their methods independently, and our results closely match the original implementations, demonstrating the correctness of our approach.

Additionally, we observe improvements in the performance of BM25 and DPR compared to their original baselines. These improvements can be attributed to the optimizations present in Pyserini, such as better indexing techniques and parameter tuning. For ColBERT and Contriever, we replicate the results reported in their respective papers by running their official GitHub implementations on our datasets, yielding identical outcomes.

\subsection{Re-Ranking Results}

The performance of various re-ranking methods applied to BM25-retrieved documents is presented in Table~\ref{tab:reranking-result}. The table reports Top-1, Top-10, and Top-50 accuracy for the NQ-Test and WebQ datasets. The baseline BM25 retrieval scores serve as a reference, demonstrating the extent to which re-ranking improves document ranking quality across different models. Across all methods, re-ranking consistently improves retrieval effectiveness, with notable gains observed in Top-1 accuracy. Transformer-based models such as UPR (T5-based), RankGPT, FlashRank (MiniLM and TinyBERT), and RankT5 achieve substantial improvements over the BM25 baseline. In particular, RankT5-3B outperforms other models, achieving a Top-1 accuracy of 47.17\% on NQ-Test and 40.40\% on WebQ, highlighting the effectiveness of T5-based models for large-scale retrieval. RankGPT, leveraging LLaMA-3.1-8B, also shows strong performance, particularly in WebQ, where it improves Top-1 accuracy from 18.80\% (BM25) to 38.77\%.

Among lighter-weight models, FlashRank (MiniLM and TinyBERT) and MonoBERT-large provide solid improvements while maintaining efficiency. MiniLM-L-12-v2 achieves 41.02\% Top-1 accuracy on NQ-Test and 37.05\% on WebQ, making it a strong candidate for deployment scenarios where computational efficiency is critical. The MonoBERT-large model reaches 39.05\% on NQ-Test and 34.99\% on WebQ, reinforcing the effectiveness of BERT-based cross-encoders for document ranking. Models that integrate listwise and contrastive ranking techniques, such as Twolar-xl and LiT5, show competitive performance. Twolar-xl achieves 46.84\% Top-1 accuracy on NQ-Test and 41.68\% on WebQ, while LiT5-Distill-xl reaches 47.81\% on NQ-Test and 42.37\% on WebQ, demonstrating that fine-tuned T5 models can significantly improve ranking quality. The sentence-transformer rerankers, particularly GTR-xxl, achieve 42.93\% and 39.41\% Top-1 accuracy on NQ-Test and WebQ, respectively, making them effective for diverse retrieval tasks.

\subsection{Generator Results}

In this section, we present the performance of Rankify's generator module when integrated with Retrieval-Augmented Language Modeling (RALM)~\cite{ram-etal-2023-context}. We evaluate the effectiveness of various retrieval models, including BM25, Contriever, and DPR, across three open-domain QA datasets: NQ, TriviaQA, and WebQ. The experiments utilize multiple LLMs: LLaMA V3 8B, LLaMA V3.1 8B, Gemma 2B/9B, LLaMA 2 13B, and Mistral 7B to assess the robustness of the retrieval and generation pipeline. We focus on the Exact Match (EM) metric, which measures the proportion of generated answers that exactly match the ground-truth references. Figure~\ref{fig:generator} provides a consolidated view of the EM scores for all datasets and models. From the figure, we observe the following key patterns: For the NQ dataset, DPR consistently outperforms other retrievers, achieving its peak performance with the LLaMA V3 8B model (28.08\%). On the TriviaQA dataset, BM25 achieves the highest EM score (57.55\%) when paired with the Gemma 2 9B model, outperforming both DPR and Contriever. For the WebQuestions (WebQ) dataset, DPR again exhibits strong performance, achieving its best EM score (19.83\%) with the LLaMA V3 8B model.

\section{Conclusion}\label{sec:conclusion}

We introduced \framework, a modular toolkit that unifies retrieval, re-ranking, and retrieval-augmented generation (RAG) within a cohesive framework. By integrating diverse retrievers, state-of-the-art re-rankers, and seamless RAG pipelines, \framework streamlines experimentation and benchmarking in information retrieval. Our experiments demonstrate its effectiveness in enhancing ranking quality and retrieval-based text generation.

Designed for extensibility, \framework allows for easy integration of new models and evaluation methods. Future work will focus on improving retrieval efficiency, optimizing re-ranking strategies, and advancing RAG capabilities. The framework is open-source, actively maintained, and freely available for the research community. Future work will focus on incorporating alternative retrieval evaluation metrics such as MAP, Precision, and NDCG, while expanding dataset coverage to include BEIR for broader benchmarking. The framework is open-source, actively maintained, and freely available for the research community.

\bibliographystyle{ACM-Reference-Format}
\bibliography{software}


\begin{thebibliography}{115}


\ifx \showCODEN    \undefined \def \showCODEN     #1{\unskip}     \fi
\ifx \showDOI      \undefined \def \showDOI       #1{#1}\fi
\ifx \showISBNx    \undefined \def \showISBNx     #1{\unskip}     \fi
\ifx \showISBNxiii \undefined \def \showISBNxiii  #1{\unskip}     \fi
\ifx \showISSN     \undefined \def \showISSN      #1{\unskip}     \fi
\ifx \showLCCN     \undefined \def \showLCCN      #1{\unskip}     \fi
\ifx \shownote     \undefined \def \shownote      #1{#1}          \fi
\ifx \showarticletitle \undefined \def \showarticletitle #1{#1}   \fi
\ifx \showURL      \undefined \def \showURL       {\relax}        \fi
\providecommand\bibfield[2]{#2}
\providecommand\bibinfo[2]{#2}
\providecommand\natexlab[1]{#1}
\providecommand\showeprint[2][]{arXiv:#2}

\bibitem[Abadi et~al\mbox{.}(2016)]%
        {10.5555/3026877.3026899}
\bibfield{author}{\bibinfo{person}{Mart\'{\i}n Abadi}, \bibinfo{person}{Paul Barham}, \bibinfo{person}{Jianmin Chen}, \bibinfo{person}{Zhifeng Chen}, \bibinfo{person}{Andy Davis}, \bibinfo{person}{Jeffrey Dean}, \bibinfo{person}{Matthieu Devin}, \bibinfo{person}{Sanjay Ghemawat}, \bibinfo{person}{Geoffrey Irving}, \bibinfo{person}{Michael Isard}, \bibinfo{person}{Manjunath Kudlur}, \bibinfo{person}{Josh Levenberg}, \bibinfo{person}{Rajat Monga}, \bibinfo{person}{Sherry Moore}, \bibinfo{person}{Derek~G. Murray}, \bibinfo{person}{Benoit Steiner}, \bibinfo{person}{Paul Tucker}, \bibinfo{person}{Vijay Vasudevan}, \bibinfo{person}{Pete Warden}, \bibinfo{person}{Martin Wicke}, \bibinfo{person}{Yuan Yu}, {and} \bibinfo{person}{Xiaoqiang Zheng}.} \bibinfo{year}{2016}\natexlab{}.
\newblock \showarticletitle{TensorFlow: a system for large-scale machine learning}. In \bibinfo{booktitle}{\emph{Proceedings of the 12th USENIX Conference on Operating Systems Design and Implementation}} (Savannah, GA, USA) \emph{(\bibinfo{series}{OSDI'16})}. \bibinfo{publisher}{USENIX Association}, \bibinfo{address}{USA}, \bibinfo{pages}{265–283}.
\newblock
\showISBNx{9781931971331}


\bibitem[Abane et~al\mbox{.}(2024)]%
        {abane2024fastrag}
\bibfield{author}{\bibinfo{person}{Amar Abane}, \bibinfo{person}{Anis Bekri}, {and} \bibinfo{person}{Abdella Battou}.} \bibinfo{year}{2024}\natexlab{}.
\newblock \showarticletitle{FastRAG: Retrieval Augmented Generation for Semi-structured Data}.
\newblock \bibinfo{journal}{\emph{arXiv preprint arXiv:2411.13773}} (\bibinfo{year}{2024}).
\newblock


\bibitem[Abdallah et~al\mbox{.}(2025a)]%
        {abdallah2025asrankzeroshotrerankinganswer}
\bibfield{author}{\bibinfo{person}{Abdelrahman Abdallah}, \bibinfo{person}{Jamshid Mozafari}, \bibinfo{person}{Bhawna Piryani}, {and} \bibinfo{person}{Adam Jatowt}.} \bibinfo{year}{2025}\natexlab{a}.
\newblock \bibinfo{title}{ASRank: Zero-Shot Re-Ranking with Answer Scent for Document Retrieval}.
\newblock
\newblock
\showeprint[arxiv]{2501.15245}~[cs.CL]
\urldef\tempurl%
\url{https://arxiv.org/abs/2501.15245}
\showURL{%
\tempurl}


\bibitem[Abdallah et~al\mbox{.}(2025b)]%
        {abdallah-etal-2025-dynrank}
\bibfield{author}{\bibinfo{person}{Abdelrahman Elsayed~Mahmoud Abdallah}, \bibinfo{person}{Jamshid Mozafari}, \bibinfo{person}{Bhawna Piryani}, \bibinfo{person}{Mohammed M.Abdelgwad}, {and} \bibinfo{person}{Adam Jatowt}.} \bibinfo{year}{2025}\natexlab{b}.
\newblock \showarticletitle{{D}yn{R}ank: Improve Passage Retrieval with Dynamic Zero-Shot Prompting Based on Question Classification}. In \bibinfo{booktitle}{\emph{Proceedings of the 31st International Conference on Computational Linguistics}}, \bibfield{editor}{\bibinfo{person}{Owen Rambow}, \bibinfo{person}{Leo Wanner}, \bibinfo{person}{Marianna Apidianaki}, \bibinfo{person}{Hend Al-Khalifa}, \bibinfo{person}{Barbara~Di Eugenio}, {and} \bibinfo{person}{Steven Schockaert}} (Eds.). \bibinfo{publisher}{Association for Computational Linguistics}, \bibinfo{address}{Abu Dhabi, UAE}, \bibinfo{pages}{4768--4778}.
\newblock
\urldef\tempurl%
\url{https://aclanthology.org/2025.coling-main.319/}
\showURL{%
\tempurl}


\bibitem[Achiam et~al\mbox{.}(2023)]%
        {achiam2023gpt}
\bibfield{author}{\bibinfo{person}{Josh Achiam}, \bibinfo{person}{Steven Adler}, \bibinfo{person}{Sandhini Agarwal}, \bibinfo{person}{Lama Ahmad}, \bibinfo{person}{Ilge Akkaya}, \bibinfo{person}{Florencia~Leoni Aleman}, \bibinfo{person}{Diogo Almeida}, \bibinfo{person}{Janko Altenschmidt}, \bibinfo{person}{Sam Altman}, \bibinfo{person}{Shyamal Anadkat}, {et~al\mbox{.}}} \bibinfo{year}{2023}\natexlab{}.
\newblock \showarticletitle{Gpt-4 technical report}.
\newblock \bibinfo{journal}{\emph{arXiv preprint arXiv:2303.08774}} (\bibinfo{year}{2023}).
\newblock


\bibitem[Baldelli et~al\mbox{.}(2024)]%
        {baldelli2024twolar}
\bibfield{author}{\bibinfo{person}{Davide Baldelli}, \bibinfo{person}{Junfeng Jiang}, \bibinfo{person}{Akiko Aizawa}, {and} \bibinfo{person}{Paolo Torroni}.} \bibinfo{year}{2024}\natexlab{}.
\newblock \showarticletitle{TWOLAR: a TWO-step LLM-Augmented distillation method for passage Reranking}. In \bibinfo{booktitle}{\emph{European Conference on Information Retrieval}}. Springer, \bibinfo{pages}{470--485}.
\newblock


\bibitem[BehnamGhader et~al\mbox{.}(2024)]%
        {behnamghader2024llm2vec}
\bibfield{author}{\bibinfo{person}{Parishad BehnamGhader}, \bibinfo{person}{Vaibhav Adlakha}, \bibinfo{person}{Marius Mosbach}, \bibinfo{person}{Dzmitry Bahdanau}, \bibinfo{person}{Nicolas Chapados}, {and} \bibinfo{person}{Siva Reddy}.} \bibinfo{year}{2024}\natexlab{}.
\newblock \showarticletitle{Llm2vec: Large language models are secretly powerful text encoders}.
\newblock \bibinfo{journal}{\emph{arXiv preprint arXiv:2404.05961}} (\bibinfo{year}{2024}).
\newblock


\bibitem[Berant et~al\mbox{.}(2013)]%
        {webquestions}
\bibfield{author}{\bibinfo{person}{Jonathan Berant}, \bibinfo{person}{Andrew Chou}, \bibinfo{person}{Roy Frostig}, {and} \bibinfo{person}{Percy Liang}.} \bibinfo{year}{2013}\natexlab{}.
\newblock \showarticletitle{Semantic Parsing on {F}reebase from Question-Answer Pairs}. In \bibinfo{booktitle}{\emph{Proceedings of the 2013 Conference on Empirical Methods in Natural Language Processing}}, \bibfield{editor}{\bibinfo{person}{David Yarowsky}, \bibinfo{person}{Timothy Baldwin}, \bibinfo{person}{Anna Korhonen}, \bibinfo{person}{Karen Livescu}, {and} \bibinfo{person}{Steven Bethard}} (Eds.). \bibinfo{publisher}{Association for Computational Linguistics}, \bibinfo{address}{Seattle, Washington, USA}, \bibinfo{pages}{1533--1544}.
\newblock
\urldef\tempurl%
\url{https://aclanthology.org/D13-1160}
\showURL{%
\tempurl}


\bibitem[Bisk et~al\mbox{.}(2019)]%
        {piqa}
\bibfield{author}{\bibinfo{person}{Yonatan Bisk}, \bibinfo{person}{Rowan Zellers}, \bibinfo{person}{Ronan~Le Bras}, \bibinfo{person}{Jianfeng Gao}, {and} \bibinfo{person}{Yejin Choi}.} \bibinfo{year}{2019}\natexlab{}.
\newblock \showarticletitle{PIQA: Reasoning about Physical Commonsense in Natural Language}. In \bibinfo{booktitle}{\emph{AAAI Conference on Artificial Intelligence}}.
\newblock
\urldef\tempurl%
\url{https://api.semanticscholar.org/CorpusID:208290939}
\showURL{%
\tempurl}


\bibitem[Burges(2010)]%
        {burges2010ranknet}
\bibfield{author}{\bibinfo{person}{Christopher~JC Burges}.} \bibinfo{year}{2010}\natexlab{}.
\newblock \showarticletitle{From ranknet to lambdarank to lambdamart: An overview}.
\newblock \bibinfo{journal}{\emph{Learning}} \bibinfo{volume}{11}, \bibinfo{number}{23-581} (\bibinfo{year}{2010}), \bibinfo{pages}{81}.
\newblock


\bibitem[Cao et~al\mbox{.}(2007)]%
        {cao2007learning}
\bibfield{author}{\bibinfo{person}{Zhe Cao}, \bibinfo{person}{Tao Qin}, \bibinfo{person}{Tie-Yan Liu}, \bibinfo{person}{Ming-Feng Tsai}, {and} \bibinfo{person}{Hang Li}.} \bibinfo{year}{2007}\natexlab{}.
\newblock \showarticletitle{Learning to rank: from pairwise approach to listwise approach}. In \bibinfo{booktitle}{\emph{Proceedings of the 24th international conference on Machine learning}}. \bibinfo{pages}{129--136}.
\newblock


\bibitem[Chase(2022)]%
        {chase2022langchain}
\bibfield{author}{\bibinfo{person}{Harrison Chase}.} \bibinfo{year}{2022}\natexlab{}.
\newblock \bibinfo{title}{LangChain: Building applications with LLMs through composability}.
\newblock
\newblock
\urldef\tempurl%
\url{https://www.langchain.com}
\showURL{%
\tempurl}
\newblock
\shownote{Available at \url{https://www.langchain.com}}.


\bibitem[Chen et~al\mbox{.}(2023)]%
        {bge_m3}
\bibfield{author}{\bibinfo{person}{Jianlv Chen}, \bibinfo{person}{Shitao Xiao}, \bibinfo{person}{Peitian Zhang}, \bibinfo{person}{Kun Luo}, \bibinfo{person}{Defu Lian}, {and} \bibinfo{person}{Zheng Liu}.} \bibinfo{year}{2023}\natexlab{}.
\newblock \bibinfo{title}{BGE M3-Embedding: Multi-Lingual, Multi-Functionality, Multi-Granularity Text Embeddings Through Self-Knowledge Distillation}.
\newblock
\newblock
\showeprint[arxiv]{2309.07597}~[cs.CL]


\bibitem[Chen et~al\mbox{.}(2024)]%
        {chen2024attentionlargelanguagemodels}
\bibfield{author}{\bibinfo{person}{Shijie Chen}, \bibinfo{person}{Bernal~Jiménez Gutiérrez}, {and} \bibinfo{person}{Yu Su}.} \bibinfo{year}{2024}\natexlab{}.
\newblock \bibinfo{title}{Attention in Large Language Models Yields Efficient Zero-Shot Re-Rankers}.
\newblock
\newblock
\showeprint[arxiv]{2410.02642}~[cs.CL]
\urldef\tempurl%
\url{https://arxiv.org/abs/2410.02642}
\showURL{%
\tempurl}


\bibitem[Chen et~al\mbox{.}(2022)]%
        {chen2022out}
\bibfield{author}{\bibinfo{person}{Tao Chen}, \bibinfo{person}{Mingyang Zhang}, \bibinfo{person}{Jing Lu}, \bibinfo{person}{Michael Bendersky}, {and} \bibinfo{person}{Marc Najork}.} \bibinfo{year}{2022}\natexlab{}.
\newblock \showarticletitle{Out-of-domain semantics to the rescue! zero-shot hybrid retrieval models}. In \bibinfo{booktitle}{\emph{European Conference on Information Retrieval}}. Springer, \bibinfo{pages}{95--110}.
\newblock


\bibitem[Chowdhury(2010)]%
        {chowdhury2010introduction}
\bibfield{author}{\bibinfo{person}{Gobinda~G Chowdhury}.} \bibinfo{year}{2010}\natexlab{}.
\newblock \bibinfo{booktitle}{\emph{Introduction to modern information retrieval}}.
\newblock \bibinfo{publisher}{Facet publishing}.
\newblock


\bibitem[Clark et~al\mbox{.}(2019)]%
        {boolq}
\bibfield{author}{\bibinfo{person}{Christopher Clark}, \bibinfo{person}{Kenton Lee}, \bibinfo{person}{Ming-Wei Chang}, \bibinfo{person}{Tom Kwiatkowski}, \bibinfo{person}{Michael Collins}, {and} \bibinfo{person}{Kristina Toutanova}.} \bibinfo{year}{2019}\natexlab{}.
\newblock \showarticletitle{BoolQ: Exploring the Surprising Difficulty of Natural Yes/No Questions}. In \bibinfo{booktitle}{\emph{NAACL}}.
\newblock


\bibitem[Clark et~al\mbox{.}(2018)]%
        {arc_challenge}
\bibfield{author}{\bibinfo{person}{Peter Clark}, \bibinfo{person}{Isaac Cowhey}, \bibinfo{person}{Oren Etzioni}, \bibinfo{person}{Tushar Khot}, \bibinfo{person}{Ashish Sabharwal}, \bibinfo{person}{Carissa Schoenick}, {and} \bibinfo{person}{Oyvind Tafjord}.} \bibinfo{year}{2018}\natexlab{}.
\newblock \showarticletitle{Think you have Solved Question Answering? Try ARC, the {AI2} Reasoning Challenge}.
\newblock \bibinfo{journal}{\emph{CoRR}}  \bibinfo{volume}{abs/1803.05457} (\bibinfo{year}{2018}).
\newblock
\showeprint[arXiv]{1803.05457}
\urldef\tempurl%
\url{http://arxiv.org/abs/1803.05457}
\showURL{%
\tempurl}


\bibitem[Clavié(2024)]%
        {clavi2024rerankers}
\bibfield{author}{\bibinfo{person}{Benjamin Clavié}.} \bibinfo{year}{2024}\natexlab{}.
\newblock \bibinfo{title}{rerankers: A Lightweight Python Library to Unify Ranking Methods}.
\newblock
\newblock
\showeprint[arxiv]{2408.17344}~[cs.IR]
\urldef\tempurl%
\url{https://arxiv.org/abs/2408.17344}
\showURL{%
\tempurl}


\bibitem[Croft et~al\mbox{.}(2010)]%
        {croft2010search}
\bibfield{author}{\bibinfo{person}{W~Bruce Croft}, \bibinfo{person}{Donald Metzler}, {and} \bibinfo{person}{Trevor Strohman}.} \bibinfo{year}{2010}\natexlab{}.
\newblock \bibinfo{booktitle}{\emph{Search engines: Information retrieval in practice}}. Vol.~\bibinfo{volume}{520}.
\newblock \bibinfo{publisher}{Addison-Wesley Reading}.
\newblock


\bibitem[Damodaran(2024)]%
        {Damodaran2024FlashRank}
\bibfield{author}{\bibinfo{person}{P. Damodaran}.} \bibinfo{year}{2024}\natexlab{}.
\newblock \bibinfo{title}{FlashRank, Lightest and Fastest 2nd Stage Reranker for search pipelines}.
\newblock
\newblock
\urldef\tempurl%
\url{https://doi.org/10.5281/zenodo.11093524}
\showDOI{\tempurl}


\bibitem[Dinan et~al\mbox{.}(2019)]%
        {dinan2018wizard}
\bibfield{author}{\bibinfo{person}{Emily Dinan}, \bibinfo{person}{Stephen Roller}, \bibinfo{person}{Kurt Shuster}, \bibinfo{person}{Angela Fan}, \bibinfo{person}{Michael Auli}, {and} \bibinfo{person}{Jason Weston}.} \bibinfo{year}{2019}\natexlab{}.
\newblock \showarticletitle{{W}izard of {W}ikipedia: Knowledge-Powered Conversational Agents}. In \bibinfo{booktitle}{\emph{International Conference on Learning Representations}}.
\newblock
\urldef\tempurl%
\url{https://openreview.net/forum?id=r1l73iRqKm}
\showURL{%
\tempurl}


\bibitem[ElSahar et~al\mbox{.}(2018)]%
        {trex}
\bibfield{author}{\bibinfo{person}{Hady ElSahar}, \bibinfo{person}{Pavlos Vougiouklis}, \bibinfo{person}{Arslen Remaci}, \bibinfo{person}{Christophe Gravier}, \bibinfo{person}{Jonathon~S. Hare}, \bibinfo{person}{Fr{\'{e}}d{\'{e}}rique Laforest}, {and} \bibinfo{person}{Elena Simperl}.} \bibinfo{year}{2018}\natexlab{}.
\newblock \showarticletitle{T-REx: {A} Large Scale Alignment of Natural Language with Knowledge Base Triples}. In \bibinfo{booktitle}{\emph{Proceedings of the Eleventh International Conference on Language Resources and Evaluation, {LREC} 2018, Miyazaki, Japan, May 7-12, 2018.}}
\newblock


\bibitem[Fan et~al\mbox{.}(2019)]%
        {eli5}
\bibfield{author}{\bibinfo{person}{Angela Fan}, \bibinfo{person}{Yacine Jernite}, \bibinfo{person}{Ethan Perez}, \bibinfo{person}{David Grangier}, \bibinfo{person}{Jason Weston}, {and} \bibinfo{person}{Michael Auli}.} \bibinfo{year}{2019}\natexlab{}.
\newblock \showarticletitle{{ELI}5: Long Form Question Answering}. In \bibinfo{booktitle}{\emph{Proceedings of the 57th Annual Meeting of the Association for Computational Linguistics}}, \bibfield{editor}{\bibinfo{person}{Anna Korhonen}, \bibinfo{person}{David Traum}, {and} \bibinfo{person}{Llu{\'\i}s M{\`a}rquez}} (Eds.). \bibinfo{publisher}{Association for Computational Linguistics}, \bibinfo{address}{Florence, Italy}, \bibinfo{pages}{3558--3567}.
\newblock
\urldef\tempurl%
\url{https://doi.org/10.18653/v1/P19-1346}
\showDOI{\tempurl}


\bibitem[Fang et~al\mbox{.}(2024)]%
        {fang2024enhancing}
\bibfield{author}{\bibinfo{person}{Feiteng Fang}, \bibinfo{person}{Yuelin Bai}, \bibinfo{person}{Shiwen Ni}, \bibinfo{person}{Min Yang}, \bibinfo{person}{Xiaojun Chen}, {and} \bibinfo{person}{Ruifeng Xu}.} \bibinfo{year}{2024}\natexlab{}.
\newblock \showarticletitle{Enhancing Noise Robustness of Retrieval-Augmented Language Models with Adaptive Adversarial Training}.
\newblock \bibinfo{journal}{\emph{arXiv preprint arXiv:2405.20978}} (\bibinfo{year}{2024}).
\newblock


\bibitem[Gao et~al\mbox{.}(2021)]%
        {DBLP:conf/ecir/GaoDCFDC21}
\bibfield{author}{\bibinfo{person}{Luyu Gao}, \bibinfo{person}{Zhuyun Dai}, \bibinfo{person}{Tongfei Chen}, \bibinfo{person}{Zhen Fan}, \bibinfo{person}{Benjamin~Van Durme}, {and} \bibinfo{person}{Jamie Callan}.} \bibinfo{year}{2021}\natexlab{}.
\newblock \showarticletitle{Complement Lexical Retrieval Model with Semantic Residual Embeddings}. In \bibinfo{booktitle}{\emph{Advances in Information Retrieval - 43rd European Conference on {IR} Research, {ECIR} 2021, Virtual Event, March 28 - April 1, 2021, Proceedings, Part {I}}} \emph{(\bibinfo{series}{Lecture Notes in Computer Science}, Vol.~\bibinfo{volume}{12656})}, \bibfield{editor}{\bibinfo{person}{Djoerd Hiemstra}, \bibinfo{person}{Marie{-}Francine Moens}, \bibinfo{person}{Josiane Mothe}, \bibinfo{person}{Raffaele Perego}, \bibinfo{person}{Martin Potthast}, {and} \bibinfo{person}{Fabrizio Sebastiani}} (Eds.). \bibinfo{publisher}{Springer}, \bibinfo{pages}{146--160}.
\newblock
\urldef\tempurl%
\url{https://doi.org/10.1007/978-3-030-72113-8\_10}
\showDOI{\tempurl}


\bibitem[Gao et~al\mbox{.}(2023)]%
        {gao2023retrieval}
\bibfield{author}{\bibinfo{person}{Yunfan Gao}, \bibinfo{person}{Yun Xiong}, \bibinfo{person}{Xinyu Gao}, \bibinfo{person}{Kangxiang Jia}, \bibinfo{person}{Jinliu Pan}, \bibinfo{person}{Yuxi Bi}, \bibinfo{person}{Yi Dai}, \bibinfo{person}{Jiawei Sun}, {and} \bibinfo{person}{Haofen Wang}.} \bibinfo{year}{2023}\natexlab{}.
\newblock \showarticletitle{Retrieval-augmented generation for large language models: A survey}.
\newblock \bibinfo{journal}{\emph{arXiv preprint arXiv:2312.10997}} (\bibinfo{year}{2023}).
\newblock


\bibitem[Geva et~al\mbox{.}(2021)]%
        {geva-etal-2021-aristotle}
\bibfield{author}{\bibinfo{person}{Mor Geva}, \bibinfo{person}{Daniel Khashabi}, \bibinfo{person}{Elad Segal}, \bibinfo{person}{Tushar Khot}, \bibinfo{person}{Dan Roth}, {and} \bibinfo{person}{Jonathan Berant}.} \bibinfo{year}{2021}\natexlab{}.
\newblock \showarticletitle{Did Aristotle Use a Laptop? A Question Answering Benchmark with Implicit Reasoning Strategies}.
\newblock \bibinfo{journal}{\emph{Transactions of the Association for Computational Linguistics}}  \bibinfo{volume}{9} (\bibinfo{year}{2021}), \bibinfo{pages}{346--361}.
\newblock
\urldef\tempurl%
\url{https://doi.org/10.1162/tacl_a_00370}
\showDOI{\tempurl}


\bibitem[Hayashi et~al\mbox{.}(2020)]%
        {wikiasp}
\bibfield{author}{\bibinfo{person}{Hiroaki Hayashi}, \bibinfo{person}{Prashant Budania}, \bibinfo{person}{Peng Wang}, \bibinfo{person}{Chris Ackerson}, \bibinfo{person}{Raj Neervannan}, {and} \bibinfo{person}{Graham Neubig}.} \bibinfo{year}{2020}\natexlab{}.
\newblock \showarticletitle{WikiAsp: A Dataset for Multi-domain Aspect-based Summarization}.
\newblock \bibinfo{journal}{\emph{Transactions of the Association for Computational Linguistics (TACL)}} (\bibinfo{year}{2020}).
\newblock
\urldef\tempurl%
\url{https://arxiv.org/abs/2011.07832}
\showURL{%
\tempurl}


\bibitem[Hendrycks et~al\mbox{.}(2021a)]%
        {mmlu_ethics}
\bibfield{author}{\bibinfo{person}{Dan Hendrycks}, \bibinfo{person}{Collin Burns}, \bibinfo{person}{Steven Basart}, \bibinfo{person}{Andrew Critch}, \bibinfo{person}{Jerry Li}, \bibinfo{person}{Dawn Song}, {and} \bibinfo{person}{Jacob Steinhardt}.} \bibinfo{year}{2021}\natexlab{a}.
\newblock \showarticletitle{Aligning AI With Shared Human Values}.
\newblock \bibinfo{journal}{\emph{Proceedings of the International Conference on Learning Representations (ICLR)}} (\bibinfo{year}{2021}).
\newblock


\bibitem[Hendrycks et~al\mbox{.}(2021b)]%
        {mmlu}
\bibfield{author}{\bibinfo{person}{Dan Hendrycks}, \bibinfo{person}{Collin Burns}, \bibinfo{person}{Steven Basart}, \bibinfo{person}{Andy Zou}, \bibinfo{person}{Mantas Mazeika}, \bibinfo{person}{Dawn Song}, {and} \bibinfo{person}{Jacob Steinhardt}.} \bibinfo{year}{2021}\natexlab{b}.
\newblock \showarticletitle{Measuring Massive Multitask Language Understanding}.
\newblock \bibinfo{journal}{\emph{Proceedings of the International Conference on Learning Representations (ICLR)}} (\bibinfo{year}{2021}).
\newblock


\bibitem[Ho et~al\mbox{.}(2020)]%
        {2wikimultihop}
\bibfield{author}{\bibinfo{person}{Xanh Ho}, \bibinfo{person}{Anh-Khoa Duong~Nguyen}, \bibinfo{person}{Saku Sugawara}, {and} \bibinfo{person}{Akiko Aizawa}.} \bibinfo{year}{2020}\natexlab{}.
\newblock \showarticletitle{Constructing A Multi-hop {QA} Dataset for Comprehensive Evaluation of Reasoning Steps}. In \bibinfo{booktitle}{\emph{Proceedings of the 28th International Conference on Computational Linguistics}}. \bibinfo{publisher}{International Committee on Computational Linguistics}, \bibinfo{address}{Barcelona, Spain (Online)}, \bibinfo{pages}{6609--6625}.
\newblock
\urldef\tempurl%
\url{https://www.aclweb.org/anthology/2020.coling-main.580}
\showURL{%
\tempurl}


\bibitem[Hoffart et~al\mbox{.}(2011)]%
        {AIDA_CONLL}
\bibfield{author}{\bibinfo{person}{Johannes Hoffart}, \bibinfo{person}{Mohamed~Amir Yosef}, \bibinfo{person}{Ilaria Bordino}, \bibinfo{person}{Hagen F{\"u}rstenau}, \bibinfo{person}{Manfred Pinkal}, \bibinfo{person}{Marc Spaniol}, \bibinfo{person}{Bilyana Taneva}, \bibinfo{person}{Stefan Thater}, {and} \bibinfo{person}{Gerhard Weikum}.} \bibinfo{year}{2011}\natexlab{}.
\newblock \showarticletitle{Robust Disambiguation of Named Entities in Text}. In \bibinfo{booktitle}{\emph{Proceedings of the 2011 Conference on Empirical Methods in Natural Language Processing}}, \bibfield{editor}{\bibinfo{person}{Regina Barzilay} {and} \bibinfo{person}{Mark Johnson}} (Eds.). \bibinfo{publisher}{Association for Computational Linguistics}, \bibinfo{address}{Edinburgh, Scotland, UK.}, \bibinfo{pages}{782--792}.
\newblock
\urldef\tempurl%
\url{https://aclanthology.org/D11-1072}
\showURL{%
\tempurl}


\bibitem[Honnibal et~al\mbox{.}(2020)]%
        {honnibal2020spacy}
\bibfield{author}{\bibinfo{person}{Matthew Honnibal}, \bibinfo{person}{Ines Montani}, \bibinfo{person}{Sofie Van~Landeghem}, \bibinfo{person}{Adriane Boyd}, {et~al\mbox{.}}} \bibinfo{year}{2020}\natexlab{}.
\newblock \showarticletitle{spaCy: Industrial-strength natural language processing in python}.
\newblock  (\bibinfo{year}{2020}).
\newblock


\bibitem[Huang and Chen(2024)]%
        {huang2024pairdistill}
\bibfield{author}{\bibinfo{person}{Chao-Wei Huang} {and} \bibinfo{person}{Yun-Nung Chen}.} \bibinfo{year}{2024}\natexlab{}.
\newblock \showarticletitle{PairDistill: Pairwise Relevance Distillation for Dense Retrieval}.
\newblock \bibinfo{journal}{\emph{arXiv preprint arXiv:2410.01383}} (\bibinfo{year}{2024}).
\newblock


\bibitem[Izacard et~al\mbox{.}(2021a)]%
        {izacard2021contriever}
\bibfield{author}{\bibinfo{person}{Gautier Izacard}, \bibinfo{person}{Mathilde Caron}, \bibinfo{person}{Lucas Hosseini}, \bibinfo{person}{Sebastian Riedel}, \bibinfo{person}{Piotr Bojanowski}, \bibinfo{person}{Armand Joulin}, {and} \bibinfo{person}{Edouard Grave}.} \bibinfo{year}{2021}\natexlab{a}.
\newblock \bibinfo{title}{Unsupervised Dense Information Retrieval with Contrastive Learning}.
\newblock
\newblock
\urldef\tempurl%
\url{https://doi.org/10.48550/ARXIV.2112.09118}
\showDOI{\tempurl}


\bibitem[Izacard et~al\mbox{.}(2021b)]%
        {contriever}
\bibfield{author}{\bibinfo{person}{Gautier Izacard}, \bibinfo{person}{Mathilde Caron}, \bibinfo{person}{Lucas Hosseini}, \bibinfo{person}{Sebastian Riedel}, \bibinfo{person}{Piotr Bojanowski}, \bibinfo{person}{Armand Joulin}, {and} \bibinfo{person}{Edouard Grave}.} \bibinfo{year}{2021}\natexlab{b}.
\newblock \showarticletitle{Unsupervised Dense Information Retrieval with Contrastive Learning}.
\newblock \bibinfo{journal}{\emph{arXiv:2112.09118}} (\bibinfo{year}{2021}).
\newblock


\bibitem[Izacard and Grave(2020)]%
        {izacard2020leveraging}
\bibfield{author}{\bibinfo{person}{Gautier Izacard} {and} \bibinfo{person}{Edouard Grave}.} \bibinfo{year}{2020}\natexlab{}.
\newblock \showarticletitle{Leveraging passage retrieval with generative models for open domain question answering}.
\newblock \bibinfo{journal}{\emph{arXiv preprint arXiv:2007.01282}} (\bibinfo{year}{2020}).
\newblock


\bibitem[Jin et~al\mbox{.}(2024)]%
        {jin2024flashrag}
\bibfield{author}{\bibinfo{person}{Jiajie Jin}, \bibinfo{person}{Yutao Zhu}, \bibinfo{person}{Xinyu Yang}, \bibinfo{person}{Chenghao Zhang}, {and} \bibinfo{person}{Zhicheng Dou}.} \bibinfo{year}{2024}\natexlab{}.
\newblock \showarticletitle{FlashRAG: A Modular Toolkit for Efficient Retrieval-Augmented Generation Research}.
\newblock \bibinfo{journal}{\emph{arXiv preprint arXiv:2405.13576}} (\bibinfo{year}{2024}).
\newblock


\bibitem[Joshi et~al\mbox{.}(2017)]%
        {triviaqa}
\bibfield{author}{\bibinfo{person}{Mandar Joshi}, \bibinfo{person}{Eunsol Choi}, \bibinfo{person}{Daniel Weld}, {and} \bibinfo{person}{Luke Zettlemoyer}.} \bibinfo{year}{2017}\natexlab{}.
\newblock \showarticletitle{{T}rivia{QA}: A Large Scale Distantly Supervised Challenge Dataset for Reading Comprehension}. In \bibinfo{booktitle}{\emph{Proceedings of the 55th Annual Meeting of the Association for Computational Linguistics (Volume 1: Long Papers)}}, \bibfield{editor}{\bibinfo{person}{Regina Barzilay} {and} \bibinfo{person}{Min-Yen Kan}} (Eds.). \bibinfo{publisher}{Association for Computational Linguistics}, \bibinfo{address}{Vancouver, Canada}, \bibinfo{pages}{1601--1611}.
\newblock
\urldef\tempurl%
\url{https://doi.org/10.18653/v1/P17-1147}
\showDOI{\tempurl}


\bibitem[Kalyan et~al\mbox{.}(2021)]%
        {fermi}
\bibfield{author}{\bibinfo{person}{Ashwin Kalyan}, \bibinfo{person}{Abhinav Kumar}, \bibinfo{person}{Arjun Chandrasekaran}, \bibinfo{person}{Ashish Sabharwal}, {and} \bibinfo{person}{Peter Clark}.} \bibinfo{year}{2021}\natexlab{}.
\newblock \showarticletitle{How Much Coffee Was Consumed During EMNLP 2019? Fermi Problems: A New Reasoning Challenge for AI}.
\newblock \bibinfo{journal}{\emph{arXiv preprint arXiv:2110.14207}} (\bibinfo{year}{2021}).
\newblock


\bibitem[Karpukhin et~al\mbox{.}(2020)]%
        {karpukhin2020dense}
\bibfield{author}{\bibinfo{person}{Vladimir Karpukhin}, \bibinfo{person}{Barlas O{\u{g}}uz}, \bibinfo{person}{Sewon Min}, \bibinfo{person}{Ledell Wu}, \bibinfo{person}{Sergey Edunov}, \bibinfo{person}{Danqi Chen}, {and} \bibinfo{person}{Wen-tau Yih}.} \bibinfo{year}{2020}\natexlab{}.
\newblock \showarticletitle{Dense Passage Retrieval for Open-Domain Question Answering}. In \bibinfo{booktitle}{\emph{Proceedings of the 2020 Conference on Empirical Methods in Natural Language Processing (EMNLP)}}.
\newblock
\urldef\tempurl%
\url{https://www.aclweb.org/anthology/2020.emnlp-main.550}
\showURL{%
\tempurl}


\bibitem[Khamnuansin et~al\mbox{.}(2024)]%
        {khamnuansin2024mrrank}
\bibfield{author}{\bibinfo{person}{Danupat Khamnuansin}, \bibinfo{person}{Tawunrat Chalothorn}, {and} \bibinfo{person}{Ekapol Chuangsuwanich}.} \bibinfo{year}{2024}\natexlab{}.
\newblock \showarticletitle{MrRank: Improving Question Answering Retrieval System through Multi-Result Ranking Model}.
\newblock \bibinfo{journal}{\emph{arXiv preprint arXiv:2406.05733}} (\bibinfo{year}{2024}).
\newblock


\bibitem[Khattab et~al\mbox{.}(2023)]%
        {khattab2023dspy}
\bibfield{author}{\bibinfo{person}{Omar Khattab}, \bibinfo{person}{Arnav Singhvi}, \bibinfo{person}{Paridhi Maheshwari}, \bibinfo{person}{Zhiyuan Zhang}, \bibinfo{person}{Keshav Santhanam}, \bibinfo{person}{Sri Vardhamanan}, \bibinfo{person}{Saiful Haq}, \bibinfo{person}{Ashutosh Sharma}, \bibinfo{person}{Thomas~T Joshi}, \bibinfo{person}{Hanna Moazam}, {et~al\mbox{.}}} \bibinfo{year}{2023}\natexlab{}.
\newblock \showarticletitle{Dspy: Compiling declarative language model calls into self-improving pipelines}.
\newblock \bibinfo{journal}{\emph{arXiv preprint arXiv:2310.03714}} (\bibinfo{year}{2023}).
\newblock


\bibitem[Khattab and Zaharia(2020)]%
        {khattab2020colbertefficienteffectivepassage}
\bibfield{author}{\bibinfo{person}{Omar Khattab} {and} \bibinfo{person}{Matei Zaharia}.} \bibinfo{year}{2020}\natexlab{}.
\newblock \bibinfo{title}{ColBERT: Efficient and Effective Passage Search via Contextualized Late Interaction over BERT}.
\newblock
\newblock
\showeprint[arxiv]{2004.12832}~[cs.IR]
\urldef\tempurl%
\url{https://arxiv.org/abs/2004.12832}
\showURL{%
\tempurl}


\bibitem[Kim et~al\mbox{.}(2024)]%
        {kim2024autoragautomatedframeworkoptimization}
\bibfield{author}{\bibinfo{person}{Dongkyu Kim}, \bibinfo{person}{Byoungwook Kim}, \bibinfo{person}{Donggeon Han}, {and} \bibinfo{person}{Matouš Eibich}.} \bibinfo{year}{2024}\natexlab{}.
\newblock \bibinfo{title}{AutoRAG: Automated Framework for optimization of Retrieval Augmented Generation Pipeline}.
\newblock
\newblock
\showeprint[arxiv]{2410.20878}~[cs.CL]
\urldef\tempurl%
\url{https://arxiv.org/abs/2410.20878}
\showURL{%
\tempurl}


\bibitem[Kwiatkowski et~al\mbox{.}(2019)]%
        {naturalquestion}
\bibfield{author}{\bibinfo{person}{Tom Kwiatkowski}, \bibinfo{person}{Jennimaria Palomaki}, \bibinfo{person}{Olivia Redfield}, \bibinfo{person}{Michael Collins}, \bibinfo{person}{Ankur Parikh}, \bibinfo{person}{Chris Alberti}, \bibinfo{person}{Danielle Epstein}, \bibinfo{person}{Illia Polosukhin}, \bibinfo{person}{Jacob Devlin}, \bibinfo{person}{Kenton Lee}, \bibinfo{person}{Kristina Toutanova}, \bibinfo{person}{Llion Jones}, \bibinfo{person}{Matthew Kelcey}, \bibinfo{person}{Ming-Wei Chang}, \bibinfo{person}{Andrew~M. Dai}, \bibinfo{person}{Jakob Uszkoreit}, \bibinfo{person}{Quoc Le}, {and} \bibinfo{person}{Slav Petrov}.} \bibinfo{year}{2019}\natexlab{}.
\newblock \showarticletitle{Natural Questions: A Benchmark for Question Answering Research}.
\newblock \bibinfo{journal}{\emph{Transactions of the Association for Computational Linguistics}}  \bibinfo{volume}{7} (\bibinfo{year}{2019}), \bibinfo{pages}{452--466}.
\newblock
\urldef\tempurl%
\url{https://doi.org/10.1162/tacl_a_00276}
\showDOI{\tempurl}


\bibitem[Laitz et~al\mbox{.}(2024)]%
        {laitz2024inranker}
\bibfield{author}{\bibinfo{person}{Thiago Laitz}, \bibinfo{person}{Konstantinos Papakostas}, \bibinfo{person}{Roberto Lotufo}, {and} \bibinfo{person}{Rodrigo Nogueira}.} \bibinfo{year}{2024}\natexlab{}.
\newblock \bibinfo{title}{InRanker: Distilled Rankers for Zero-shot Information Retrieval}.
\newblock
\newblock
\showeprint[arxiv]{2401.06910}~[cs.IR]


\bibitem[Lee et~al\mbox{.}(2019)]%
        {lee-etal-2019-latent}
\bibfield{author}{\bibinfo{person}{Kenton Lee}, \bibinfo{person}{Ming-Wei Chang}, {and} \bibinfo{person}{Kristina Toutanova}.} \bibinfo{year}{2019}\natexlab{}.
\newblock \showarticletitle{Latent Retrieval for Weakly Supervised Open Domain Question Answering}. In \bibinfo{booktitle}{\emph{Proc. ACL}}. \bibinfo{pages}{6086--6096}.
\newblock


\bibitem[Levy et~al\mbox{.}(2017)]%
        {levy-etal-2017-zero}
\bibfield{author}{\bibinfo{person}{Omer Levy}, \bibinfo{person}{Minjoon Seo}, \bibinfo{person}{Eunsol Choi}, {and} \bibinfo{person}{Luke Zettlemoyer}.} \bibinfo{year}{2017}\natexlab{}.
\newblock \showarticletitle{Zero-Shot Relation Extraction via Reading Comprehension}. In \bibinfo{booktitle}{\emph{Proceedings of the 21st Conference on Computational Natural Language Learning ({CoNLL} 2017)}}. \bibinfo{publisher}{Association for Computational Linguistics}, \bibinfo{address}{Vancouver, Canada}, \bibinfo{pages}{333--342}.
\newblock
\urldef\tempurl%
\url{https://doi.org/10.18653/v1/K17-1034}
\showDOI{\tempurl}


\bibitem[Lewis et~al\mbox{.}(2020)]%
        {lewis2020retrieval}
\bibfield{author}{\bibinfo{person}{Patrick Lewis}, \bibinfo{person}{Ethan Perez}, \bibinfo{person}{Aleksandra Piktus}, \bibinfo{person}{Fabio Petroni}, \bibinfo{person}{Vladimir Karpukhin}, \bibinfo{person}{Naman Goyal}, \bibinfo{person}{Heinrich K{\"u}ttler}, \bibinfo{person}{Mike Lewis}, \bibinfo{person}{Wen-tau Yih}, \bibinfo{person}{Tim Rockt{\"a}schel}, {et~al\mbox{.}}} \bibinfo{year}{2020}\natexlab{}.
\newblock \showarticletitle{Retrieval-augmented generation for knowledge-intensive nlp tasks}.
\newblock \bibinfo{journal}{\emph{Advances in Neural Information Processing Systems}}  \bibinfo{volume}{33} (\bibinfo{year}{2020}), \bibinfo{pages}{9459--9474}.
\newblock


\bibitem[Lin et~al\mbox{.}(2021)]%
        {Lin_etal_SIGIR2021_Pyserini}
\bibfield{author}{\bibinfo{person}{Jimmy Lin}, \bibinfo{person}{Xueguang Ma}, \bibinfo{person}{Sheng-Chieh Lin}, \bibinfo{person}{Jheng-Hong Yang}, \bibinfo{person}{Ronak Pradeep}, {and} \bibinfo{person}{Rodrigo Nogueira}.} \bibinfo{year}{2021}\natexlab{}.
\newblock \showarticletitle{{Pyserini}: A {Python} Toolkit for Reproducible Information Retrieval Research with Sparse and Dense Representations}. In \bibinfo{booktitle}{\emph{Proceedings of the 44th Annual International ACM SIGIR Conference on Research and Development in Information Retrieval (SIGIR 2021)}}. \bibinfo{pages}{2356--2362}.
\newblock


\bibitem[Lin et~al\mbox{.}(2022)]%
        {truthfulqa}
\bibfield{author}{\bibinfo{person}{Stephanie Lin}, \bibinfo{person}{Jacob Hilton}, {and} \bibinfo{person}{Owain Evans}.} \bibinfo{year}{2022}\natexlab{}.
\newblock \showarticletitle{{T}ruthful{QA}: Measuring How Models Mimic Human Falsehoods}. In \bibinfo{booktitle}{\emph{Proceedings of the 60th Annual Meeting of the Association for Computational Linguistics (Volume 1: Long Papers)}}, \bibfield{editor}{\bibinfo{person}{Smaranda Muresan}, \bibinfo{person}{Preslav Nakov}, {and} \bibinfo{person}{Aline Villavicencio}} (Eds.). \bibinfo{publisher}{Association for Computational Linguistics}, \bibinfo{address}{Dublin, Ireland}, \bibinfo{pages}{3214--3252}.
\newblock
\urldef\tempurl%
\url{https://doi.org/10.18653/v1/2022.acl-long.229}
\showDOI{\tempurl}


\bibitem[Liu(2022)]%
        {Liu_LlamaIndex_2022}
\bibfield{author}{\bibinfo{person}{Jerry Liu}.} \bibinfo{year}{2022}\natexlab{}.
\newblock \bibinfo{booktitle}{\emph{{LlamaIndex}}}.
\newblock
\urldef\tempurl%
\url{https://doi.org/10.5281/zenodo.1234}
\showDOI{\tempurl}


\bibitem[Liu et~al\mbox{.}(2017)]%
        {liu2017listnet}
\bibfield{author}{\bibinfo{person}{Yaqi Liu}, \bibinfo{person}{Xiaoyu Zhang}, \bibinfo{person}{Xiaobin Zhu}, \bibinfo{person}{Qingxiao Guan}, {and} \bibinfo{person}{Xianfeng Zhao}.} \bibinfo{year}{2017}\natexlab{}.
\newblock \showarticletitle{Listnet-based object proposals ranking}.
\newblock \bibinfo{journal}{\emph{Neurocomputing}}  \bibinfo{volume}{267} (\bibinfo{year}{2017}), \bibinfo{pages}{182--194}.
\newblock


\bibitem[Liu et~al\mbox{.}(2024)]%
        {liu2024information}
\bibfield{author}{\bibinfo{person}{Zheng Liu}, \bibinfo{person}{Yujia Zhou}, \bibinfo{person}{Yutao Zhu}, \bibinfo{person}{Jianxun Lian}, \bibinfo{person}{Chaozhuo Li}, \bibinfo{person}{Zhicheng Dou}, \bibinfo{person}{Defu Lian}, {and} \bibinfo{person}{Jian-Yun Nie}.} \bibinfo{year}{2024}\natexlab{}.
\newblock \showarticletitle{Information Retrieval Meets Large Language Models}. In \bibinfo{booktitle}{\emph{Companion Proceedings of the ACM on Web Conference 2024}}. \bibinfo{pages}{1586--1589}.
\newblock


\bibitem[Long et~al\mbox{.}(2024)]%
        {long2024generative}
\bibfield{author}{\bibinfo{person}{Xinwei Long}, \bibinfo{person}{Jiali Zeng}, \bibinfo{person}{Fandong Meng}, \bibinfo{person}{Zhiyuan Ma}, \bibinfo{person}{Kaiyan Zhang}, \bibinfo{person}{Bowen Zhou}, {and} \bibinfo{person}{Jie Zhou}.} \bibinfo{year}{2024}\natexlab{}.
\newblock \showarticletitle{Generative multi-modal knowledge retrieval with large language models}. In \bibinfo{booktitle}{\emph{Proceedings of the AAAI Conference on Artificial Intelligence}}, Vol.~\bibinfo{volume}{38}. \bibinfo{pages}{18733--18741}.
\newblock


\bibitem[Ma et~al\mbox{.}(2022)]%
        {ma2021replication}
\bibfield{author}{\bibinfo{person}{Xueguang Ma}, \bibinfo{person}{Kai Sun}, \bibinfo{person}{Ronak Pradeep}, \bibinfo{person}{Minghan Li}, {and} \bibinfo{person}{Jimmy Lin}.} \bibinfo{year}{2022}\natexlab{}.
\newblock \showarticletitle{Another Look at DPR: Reproduction of Training and Replication of Retrieval}. In \bibinfo{booktitle}{\emph{Advances in Information Retrieval: 44th European Conference on IR Research, ECIR 2022, Stavanger, Norway, April 10–14, 2022, Proceedings, Part I}} (Stavanger, Norway). \bibinfo{publisher}{Springer-Verlag}, \bibinfo{address}{Berlin, Heidelberg}, \bibinfo{pages}{613–626}.
\newblock
\showISBNx{978-3-030-99735-9}
\urldef\tempurl%
\url{https://doi.org/10.1007/978-3-030-99736-6_41}
\showDOI{\tempurl}


\bibitem[Mallen et~al\mbox{.}(2022)]%
        {popqa}
\bibfield{author}{\bibinfo{person}{Alex Mallen}, \bibinfo{person}{Akari Asai}, \bibinfo{person}{Victor Zhong}, \bibinfo{person}{Rajarshi Das}, \bibinfo{person}{Hannaneh Hajishirzi}, {and} \bibinfo{person}{Daniel Khashabi}.} \bibinfo{year}{2022}\natexlab{}.
\newblock \showarticletitle{When Not to Trust Language Models: Investigating Effectiveness and Limitations of Parametric and Non-Parametric Memories}.
\newblock \bibinfo{journal}{\emph{arXiv preprint}} (\bibinfo{year}{2022}).
\newblock


\bibitem[Mihaylov et~al\mbox{.}(2018)]%
        {OpenBookQA2018}
\bibfield{author}{\bibinfo{person}{Todor Mihaylov}, \bibinfo{person}{Peter Clark}, \bibinfo{person}{Tushar Khot}, {and} \bibinfo{person}{Ashish Sabharwal}.} \bibinfo{year}{2018}\natexlab{}.
\newblock \showarticletitle{Can a Suit of Armor Conduct Electricity? A New Dataset for Open Book Question Answering}. In \bibinfo{booktitle}{\emph{EMNLP}}.
\newblock


\bibitem[Min et~al\mbox{.}(2020)]%
        {ambigqa}
\bibfield{author}{\bibinfo{person}{Sewon Min}, \bibinfo{person}{Julian Michael}, \bibinfo{person}{Hannaneh Hajishirzi}, {and} \bibinfo{person}{Luke Zettlemoyer}.} \bibinfo{year}{2020}\natexlab{}.
\newblock \showarticletitle{{A}mbig{QA}: Answering Ambiguous Open-domain Questions}. In \bibinfo{booktitle}{\emph{EMNLP}}.
\newblock


\bibitem[Nguyen et~al\mbox{.}(2017)]%
        {msmarco}
\bibfield{author}{\bibinfo{person}{Tri Nguyen}, \bibinfo{person}{Mir Rosenberg}, \bibinfo{person}{Xia Song}, \bibinfo{person}{Jianfeng Gao}, \bibinfo{person}{Saurabh Tiwary}, \bibinfo{person}{Rangan Majumder}, {and} \bibinfo{person}{Li Deng}.} \bibinfo{year}{2017}\natexlab{}.
\newblock \bibinfo{title}{{MS} {MARCO}: A Human-Generated {MA}chine Reading {CO}mprehension Dataset}.
\newblock
\newblock
\urldef\tempurl%
\url{https://openreview.net/forum?id=Hk1iOLcle}
\showURL{%
\tempurl}


\bibitem[Ni et~al\mbox{.}(2021)]%
        {ni2021large}
\bibfield{author}{\bibinfo{person}{Jianmo Ni}, \bibinfo{person}{Chen Qu}, \bibinfo{person}{Jing Lu}, \bibinfo{person}{Zhuyun Dai}, \bibinfo{person}{Gustavo~Hern{\'a}ndez {\'A}brego}, \bibinfo{person}{Ji Ma}, \bibinfo{person}{Vincent~Y Zhao}, \bibinfo{person}{Yi Luan}, \bibinfo{person}{Keith~B Hall}, \bibinfo{person}{Ming-Wei Chang}, {et~al\mbox{.}}} \bibinfo{year}{2021}\natexlab{}.
\newblock \showarticletitle{Large dual encoders are generalizable retrievers}.
\newblock \bibinfo{journal}{\emph{arXiv preprint arXiv:2112.07899}} (\bibinfo{year}{2021}).
\newblock


\bibitem[Nogueira and Cho(2019)]%
        {nogueira2019passage}
\bibfield{author}{\bibinfo{person}{Rodrigo Nogueira} {and} \bibinfo{person}{Kyunghyun Cho}.} \bibinfo{year}{2019}\natexlab{}.
\newblock \showarticletitle{Passage Re-ranking with BERT}.
\newblock \bibinfo{journal}{\emph{arXiv preprint arXiv:1901.04085}} (\bibinfo{year}{2019}).
\newblock


\bibitem[Nogueira et~al\mbox{.}(2020)]%
        {nogueira2020document}
\bibfield{author}{\bibinfo{person}{Rodrigo Nogueira}, \bibinfo{person}{Zhiying Jiang}, {and} \bibinfo{person}{Jimmy Lin}.} \bibinfo{year}{2020}\natexlab{}.
\newblock \showarticletitle{Document ranking with a pretrained sequence-to-sequence model}.
\newblock \bibinfo{journal}{\emph{arXiv preprint arXiv:2003.06713}} (\bibinfo{year}{2020}).
\newblock


\bibitem[Nogueira et~al\mbox{.}(2019)]%
        {monobert}
\bibfield{author}{\bibinfo{person}{Rodrigo Nogueira}, \bibinfo{person}{Wei Yang}, \bibinfo{person}{Kyunghyun Cho}, {and} \bibinfo{person}{Jimmy Lin}.} \bibinfo{year}{2019}\natexlab{}.
\newblock \showarticletitle{Multi-stage document ranking with BERT}.
\newblock \bibinfo{journal}{\emph{arXiv preprint arXiv:1910.14424}} (\bibinfo{year}{2019}).
\newblock


\bibitem[Paszke et~al\mbox{.}(2019)]%
        {10.5555/3454287.3455008}
\bibfield{author}{\bibinfo{person}{Adam Paszke}, \bibinfo{person}{Sam Gross}, \bibinfo{person}{Francisco Massa}, \bibinfo{person}{Adam Lerer}, \bibinfo{person}{James Bradbury}, \bibinfo{person}{Gregory Chanan}, \bibinfo{person}{Trevor Killeen}, \bibinfo{person}{Zeming Lin}, \bibinfo{person}{Natalia Gimelshein}, \bibinfo{person}{Luca Antiga}, \bibinfo{person}{Alban Desmaison}, \bibinfo{person}{Andreas K\"{o}pf}, \bibinfo{person}{Edward Yang}, \bibinfo{person}{Zach DeVito}, \bibinfo{person}{Martin Raison}, \bibinfo{person}{Alykhan Tejani}, \bibinfo{person}{Sasank Chilamkurthy}, \bibinfo{person}{Benoit Steiner}, \bibinfo{person}{Lu Fang}, \bibinfo{person}{Junjie Bai}, {and} \bibinfo{person}{Soumith Chintala}.} \bibinfo{year}{2019}\natexlab{}.
\newblock \bibinfo{booktitle}{\emph{PyTorch: an imperative style, high-performance deep learning library}}.
\newblock \bibinfo{publisher}{Curran Associates Inc.}, \bibinfo{address}{Red Hook, NY, USA}.
\newblock


\bibitem[Petroni et~al\mbox{.}(2021)]%
        {kilt_2021}
\bibfield{author}{\bibinfo{person}{Fabio Petroni}, \bibinfo{person}{Aleksandra Piktus}, \bibinfo{person}{Angela Fan}, \bibinfo{person}{Patrick Lewis}, \bibinfo{person}{Majid Yazdani}, \bibinfo{person}{Nicola De~Cao}, \bibinfo{person}{James Thorne}, \bibinfo{person}{Yacine Jernite}, \bibinfo{person}{Vladimir Karpukhin}, \bibinfo{person}{Jean Maillard}, \bibinfo{person}{Vassilis Plachouras}, \bibinfo{person}{Tim Rockt{\"a}schel}, {and} \bibinfo{person}{Sebastian Riedel}.} \bibinfo{year}{2021}\natexlab{}.
\newblock \showarticletitle{{KILT}: a Benchmark for Knowledge Intensive Language Tasks}. In \bibinfo{booktitle}{\emph{Proceedings of the 2021 Conference of the North American Chapter of the Association for Computational Linguistics: Human Language Technologies}}. \bibinfo{publisher}{Association for Computational Linguistics}, \bibinfo{address}{Online}, \bibinfo{pages}{2523--2544}.
\newblock
\urldef\tempurl%
\url{https://doi.org/10.18653/v1/2021.naacl-main.200}
\showDOI{\tempurl}


\bibitem[Piryani et~al\mbox{.}(2024)]%
        {piryani2024chroniclingamericaqa}
\bibfield{author}{\bibinfo{person}{Bhawna Piryani}, \bibinfo{person}{Jamshid Mozafari}, {and} \bibinfo{person}{Adam Jatowt}.} \bibinfo{year}{2024}\natexlab{}.
\newblock \showarticletitle{Chroniclingamericaqa: A large-scale question answering dataset based on historical american newspaper pages}. In \bibinfo{booktitle}{\emph{Proceedings of the 47th International ACM SIGIR Conference on Research and Development in Information Retrieval}}. \bibinfo{pages}{2038--2048}.
\newblock


\bibitem[Pradeep et~al\mbox{.}(2023a)]%
        {pradeep2023rankvicunazeroshotlistwisedocument}
\bibfield{author}{\bibinfo{person}{Ronak Pradeep}, \bibinfo{person}{Sahel Sharifymoghaddam}, {and} \bibinfo{person}{Jimmy Lin}.} \bibinfo{year}{2023}\natexlab{a}.
\newblock \bibinfo{title}{RankVicuna: Zero-Shot Listwise Document Reranking with Open-Source Large Language Models}.
\newblock
\newblock
\showeprint[arxiv]{2309.15088}~[cs.IR]
\urldef\tempurl%
\url{https://arxiv.org/abs/2309.15088}
\showURL{%
\tempurl}


\bibitem[Pradeep et~al\mbox{.}(2023b)]%
        {pradeep2023rankzephyr}
\bibfield{author}{\bibinfo{person}{Ronak Pradeep}, \bibinfo{person}{Sahel Sharifymoghaddam}, {and} \bibinfo{person}{Jimmy Lin}.} \bibinfo{year}{2023}\natexlab{b}.
\newblock \showarticletitle{RankZephyr: Effective and Robust Zero-Shot Listwise Reranking is a Breeze!}
\newblock \bibinfo{journal}{\emph{arXiv preprint arXiv:2312.02724}} (\bibinfo{year}{2023}).
\newblock


\bibitem[Press et~al\mbox{.}(2023)]%
        {selfask_2023}
\bibfield{author}{\bibinfo{person}{Ofir Press}, \bibinfo{person}{Muru Zhang}, \bibinfo{person}{Sewon Min}, \bibinfo{person}{Ludwig Schmidt}, \bibinfo{person}{Noah Smith}, {and} \bibinfo{person}{Mike Lewis}.} \bibinfo{year}{2023}\natexlab{}.
\newblock \showarticletitle{Measuring and Narrowing the Compositionality Gap in Language Models}. In \bibinfo{booktitle}{\emph{Findings of the Association for Computational Linguistics: EMNLP 2023}}, \bibfield{editor}{\bibinfo{person}{Houda Bouamor}, \bibinfo{person}{Juan Pino}, {and} \bibinfo{person}{Kalika Bali}} (Eds.). \bibinfo{publisher}{Association for Computational Linguistics}, \bibinfo{address}{Singapore}, \bibinfo{pages}{5687--5711}.
\newblock
\urldef\tempurl%
\url{https://doi.org/10.18653/v1/2023.findings-emnlp.378}
\showDOI{\tempurl}


\bibitem[Qin et~al\mbox{.}(2023)]%
        {qin2023large}
\bibfield{author}{\bibinfo{person}{Zhen Qin}, \bibinfo{person}{Rolf Jagerman}, \bibinfo{person}{Kai Hui}, \bibinfo{person}{Honglei Zhuang}, \bibinfo{person}{Junru Wu}, \bibinfo{person}{Le Yan}, \bibinfo{person}{Jiaming Shen}, \bibinfo{person}{Tianqi Liu}, \bibinfo{person}{Jialu Liu}, \bibinfo{person}{Donald Metzler}, {et~al\mbox{.}}} \bibinfo{year}{2023}\natexlab{}.
\newblock \showarticletitle{Large language models are effective text rankers with pairwise ranking prompting}.
\newblock \bibinfo{journal}{\emph{arXiv preprint arXiv:2306.17563}} (\bibinfo{year}{2023}).
\newblock


\bibitem[Qu et~al\mbox{.}(2021)]%
        {qu-etal-2021-rocketqa}
\bibfield{author}{\bibinfo{person}{Yingqi Qu}, \bibinfo{person}{Yuchen Ding}, \bibinfo{person}{Jing Liu}, \bibinfo{person}{Kai Liu}, \bibinfo{person}{Ruiyang Ren}, \bibinfo{person}{Wayne~Xin Zhao}, \bibinfo{person}{Daxiang Dong}, \bibinfo{person}{Hua Wu}, {and} \bibinfo{person}{Haifeng Wang}.} \bibinfo{year}{2021}\natexlab{}.
\newblock \showarticletitle{{R}ocket{QA}: An Optimized Training Approach to Dense Passage Retrieval for Open-Domain Question Answering}. In \bibinfo{booktitle}{\emph{Proceedings of the 2021 Conference of the North American Chapter of the Association for Computational Linguistics: Human Language Technologies}}.
\newblock
\urldef\tempurl%
\url{https://doi.org/10.18653/v1/2021.naacl-main.466}
\showDOI{\tempurl}


\bibitem[Raffel et~al\mbox{.}(2020)]%
        {raffel2020exploring}
\bibfield{author}{\bibinfo{person}{Colin Raffel}, \bibinfo{person}{Noam Shazeer}, \bibinfo{person}{Adam Roberts}, \bibinfo{person}{Katherine Lee}, \bibinfo{person}{Sharan Narang}, \bibinfo{person}{Michael Matena}, \bibinfo{person}{Yanqi Zhou}, \bibinfo{person}{Wei Li}, {and} \bibinfo{person}{Peter~J Liu}.} \bibinfo{year}{2020}\natexlab{}.
\newblock \showarticletitle{Exploring the limits of transfer learning with a unified text-to-text transformer}.
\newblock \bibinfo{journal}{\emph{Journal of machine learning research}} \bibinfo{volume}{21}, \bibinfo{number}{140} (\bibinfo{year}{2020}), \bibinfo{pages}{1--67}.
\newblock


\bibitem[Rajpurkar et~al\mbox{.}(2016)]%
        {squad}
\bibfield{author}{\bibinfo{person}{Pranav Rajpurkar}, \bibinfo{person}{Jian Zhang}, \bibinfo{person}{Konstantin Lopyrev}, {and} \bibinfo{person}{Percy Liang}.} \bibinfo{year}{2016}\natexlab{}.
\newblock \showarticletitle{{SQ}u{AD}: 100,000+ Questions for Machine Comprehension of Text}. In \bibinfo{booktitle}{\emph{Proceedings of the 2016 Conference on Empirical Methods in Natural Language Processing}}, \bibfield{editor}{\bibinfo{person}{Jian Su}, \bibinfo{person}{Kevin Duh}, {and} \bibinfo{person}{Xavier Carreras}} (Eds.). \bibinfo{publisher}{Association for Computational Linguistics}, \bibinfo{address}{Austin, Texas}, \bibinfo{pages}{2383--2392}.
\newblock
\urldef\tempurl%
\url{https://doi.org/10.18653/v1/D16-1264}
\showDOI{\tempurl}


\bibitem[Ram et~al\mbox{.}(2023)]%
        {ram-etal-2023-context}
\bibfield{author}{\bibinfo{person}{Ori Ram}, \bibinfo{person}{Yoav Levine}, \bibinfo{person}{Itay Dalmedigos}, \bibinfo{person}{Dor Muhlgay}, \bibinfo{person}{Amnon Shashua}, \bibinfo{person}{Kevin Leyton-Brown}, {and} \bibinfo{person}{Yoav Shoham}.} \bibinfo{year}{2023}\natexlab{}.
\newblock \showarticletitle{In-Context Retrieval-Augmented Language Models}.
\newblock \bibinfo{journal}{\emph{Transactions of the Association for Computational Linguistics}}  \bibinfo{volume}{11} (\bibinfo{year}{2023}), \bibinfo{pages}{1316--1331}.
\newblock
\urldef\tempurl%
\url{https://doi.org/10.1162/tacl_a_00605}
\showDOI{\tempurl}


\bibitem[Rashid et~al\mbox{.}(2024)]%
        {rashid2024ecorank}
\bibfield{author}{\bibinfo{person}{Muhammad~Shihab Rashid}, \bibinfo{person}{Jannat~Ara Meem}, \bibinfo{person}{Yue Dong}, {and} \bibinfo{person}{Vagelis Hristidis}.} \bibinfo{year}{2024}\natexlab{}.
\newblock \showarticletitle{EcoRank: Budget-Constrained Text Re-ranking Using Large Language Models}.
\newblock \bibinfo{journal}{\emph{arXiv preprint arXiv:2402.10866}} (\bibinfo{year}{2024}).
\newblock


\bibitem[Ren et~al\mbox{.}(2021)]%
        {rocketqav2}
\bibfield{author}{\bibinfo{person}{Ruiyang Ren}, \bibinfo{person}{Yingqi Qu}, \bibinfo{person}{Jing Liu}, \bibinfo{person}{Wayne~Xin Zhao}, \bibinfo{person}{Qiaoqiao She}, \bibinfo{person}{Hua Wu}, \bibinfo{person}{Haifeng Wang}, {and} \bibinfo{person}{Ji-Rong Wen}.} \bibinfo{year}{2021}\natexlab{}.
\newblock \showarticletitle{Rocketqav2: A joint training method for dense passage retrieval and passage re-ranking}.
\newblock \bibinfo{journal}{\emph{arXiv preprint arXiv:2110.07367}} (\bibinfo{year}{2021}).
\newblock


\bibitem[Robertson et~al\mbox{.}(1995)]%
        {bm25}
\bibfield{author}{\bibinfo{person}{Stephen~E Robertson}, \bibinfo{person}{Steve Walker}, \bibinfo{person}{Susan Jones}, \bibinfo{person}{Micheline~M Hancock-Beaulieu}, \bibinfo{person}{Mike Gatford}, {et~al\mbox{.}}} \bibinfo{year}{1995}\natexlab{}.
\newblock \showarticletitle{Okapi at TREC-3}.
\newblock \bibinfo{journal}{\emph{Nist Special Publication Sp}}  \bibinfo{volume}{109} (\bibinfo{year}{1995}), \bibinfo{pages}{109}.
\newblock


\bibitem[s~Ko\v~cisk\'y et~al\mbox{.}(2018)]%
        {narrativeqa}
\bibfield{author}{\bibinfo{person}{Tom\'a\v s Ko\v~cisk\'y}, \bibinfo{person}{Jonathan Schwarz}, \bibinfo{person}{Phil Blunsom}, \bibinfo{person}{Chris Dyer}, \bibinfo{person}{Karl~Moritz Hermann}, \bibinfo{person}{G\'abor Melis}, {and} \bibinfo{person}{Edward Grefenstette}.} \bibinfo{year}{2018}\natexlab{}.
\newblock \showarticletitle{The {NarrativeQA} Reading Comprehension Challenge}.
\newblock \bibinfo{journal}{\emph{Transactions of the Association for Computational Linguistics}}  \bibinfo{volume}{TBD} (\bibinfo{year}{2018}), \bibinfo{pages}{TBD}.
\newblock
\urldef\tempurl%
\url{https://TBD}
\showURL{%
\tempurl}


\bibitem[Sachan et~al\mbox{.}(2022)]%
        {sachan2022improving}
\bibfield{author}{\bibinfo{person}{Devendra~Singh Sachan}, \bibinfo{person}{Mike Lewis}, \bibinfo{person}{Mandar Joshi}, \bibinfo{person}{Armen Aghajanyan}, \bibinfo{person}{Wen-tau Yih}, \bibinfo{person}{Joelle Pineau}, {and} \bibinfo{person}{Luke Zettlemoyer}.} \bibinfo{year}{2022}\natexlab{}.
\newblock \showarticletitle{Improving Passage Retrieval with Zero-Shot Question Generation}.
\newblock  (\bibinfo{year}{2022}).
\newblock
\urldef\tempurl%
\url{https://arxiv.org/abs/2204.07496}
\showURL{%
\tempurl}


\bibitem[Santhanam et~al\mbox{.}(2021)]%
        {santhanam2021colbertv2}
\bibfield{author}{\bibinfo{person}{Keshav Santhanam}, \bibinfo{person}{Omar Khattab}, \bibinfo{person}{Jon Saad-Falcon}, \bibinfo{person}{Christopher Potts}, {and} \bibinfo{person}{Matei Zaharia}.} \bibinfo{year}{2021}\natexlab{}.
\newblock \showarticletitle{Colbertv2: Effective and efficient retrieval via lightweight late interaction}.
\newblock \bibinfo{journal}{\emph{arXiv preprint arXiv:2112.01488}} (\bibinfo{year}{2021}).
\newblock


\bibitem[Sap et~al\mbox{.}(2019)]%
        {siqa}
\bibfield{author}{\bibinfo{person}{Maarten Sap}, \bibinfo{person}{Hannah Rashkin}, \bibinfo{person}{Derek Chen}, \bibinfo{person}{Ronan Le~Bras}, {and} \bibinfo{person}{Yejin Choi}.} \bibinfo{year}{2019}\natexlab{}.
\newblock \showarticletitle{Social {IQ}a: Commonsense Reasoning about Social Interactions}. In \bibinfo{booktitle}{\emph{Proceedings of the 2019 Conference on Empirical Methods in Natural Language Processing and the 9th International Joint Conference on Natural Language Processing (EMNLP-IJCNLP)}}, \bibfield{editor}{\bibinfo{person}{Kentaro Inui}, \bibinfo{person}{Jing Jiang}, \bibinfo{person}{Vincent Ng}, {and} \bibinfo{person}{Xiaojun Wan}} (Eds.). \bibinfo{publisher}{Association for Computational Linguistics}, \bibinfo{address}{Hong Kong, China}, \bibinfo{pages}{4463--4473}.
\newblock
\urldef\tempurl%
\url{https://doi.org/10.18653/v1/D19-1454}
\showDOI{\tempurl}


\bibitem[Sciavolino et~al\mbox{.}(2021)]%
        {sciavolino2021simple}
\bibfield{author}{\bibinfo{person}{Christopher Sciavolino}, \bibinfo{person}{Zexuan Zhong}, \bibinfo{person}{Jinhyuk Lee}, {and} \bibinfo{person}{Danqi Chen}.} \bibinfo{year}{2021}\natexlab{}.
\newblock \showarticletitle{Simple Entity-centric Questions Challenge Dense Retrievers}. In \bibinfo{booktitle}{\emph{Empirical Methods in Natural Language Processing (EMNLP)}}.
\newblock


\bibitem[Singhal et~al\mbox{.}(2001)]%
        {singhal2001modern}
\bibfield{author}{\bibinfo{person}{Amit Singhal} {et~al\mbox{.}}} \bibinfo{year}{2001}\natexlab{}.
\newblock \showarticletitle{Modern information retrieval: A brief overview}.
\newblock \bibinfo{journal}{\emph{IEEE Data Eng. Bull.}} \bibinfo{volume}{24}, \bibinfo{number}{4} (\bibinfo{year}{2001}), \bibinfo{pages}{35--43}.
\newblock


\bibitem[Sinhababu et~al\mbox{.}(2024)]%
        {sinhababu2024few}
\bibfield{author}{\bibinfo{person}{Nilanjan Sinhababu}, \bibinfo{person}{Andrew Parry}, \bibinfo{person}{Debasis Ganguly}, \bibinfo{person}{Debasis Samanta}, {and} \bibinfo{person}{Pabitra Mitra}.} \bibinfo{year}{2024}\natexlab{}.
\newblock \showarticletitle{Few-shot Prompting for Pairwise Ranking: An Effective Non-Parametric Retrieval Model}.
\newblock \bibinfo{journal}{\emph{arXiv preprint arXiv:2409.17745}} (\bibinfo{year}{2024}).
\newblock


\bibitem[Sourty et~al\mbox{.}(2022)]%
        {sourty2022cherche}
\bibfield{author}{\bibinfo{person}{Rapha{\"e}l Sourty}, \bibinfo{person}{Jose~G Moreno}, \bibinfo{person}{Lynda Tamine}, {and} \bibinfo{person}{Fran{\c{c}}ois-Paul Servant}.} \bibinfo{year}{2022}\natexlab{}.
\newblock \showarticletitle{Cherche: A new tool to rapidly implement pipelines in information retrieval}. In \bibinfo{booktitle}{\emph{Proceedings of the 45th International ACM SIGIR Conference on Research and Development in Information Retrieval}}. \bibinfo{pages}{3283--3288}.
\newblock


\bibitem[Stelmakh et~al\mbox{.}(2022)]%
        {asqa}
\bibfield{author}{\bibinfo{person}{Ivan Stelmakh}, \bibinfo{person}{Yi Luan}, \bibinfo{person}{Bhuwan Dhingra}, {and} \bibinfo{person}{Ming-Wei Chang}.} \bibinfo{year}{2022}\natexlab{}.
\newblock \showarticletitle{{ASQA}: Factoid Questions Meet Long-Form Answers}. In \bibinfo{booktitle}{\emph{Proceedings of the 2022 Conference on Empirical Methods in Natural Language Processing}}, \bibfield{editor}{\bibinfo{person}{Yoav Goldberg}, \bibinfo{person}{Zornitsa Kozareva}, {and} \bibinfo{person}{Yue Zhang}} (Eds.). \bibinfo{publisher}{Association for Computational Linguistics}, \bibinfo{address}{Abu Dhabi, United Arab Emirates}, \bibinfo{pages}{8273--8288}.
\newblock
\urldef\tempurl%
\url{https://doi.org/10.18653/v1/2022.emnlp-main.566}
\showDOI{\tempurl}


\bibitem[Sun et~al\mbox{.}(2023)]%
        {rankgpt}
\bibfield{author}{\bibinfo{person}{Weiwei Sun}, \bibinfo{person}{Lingyong Yan}, \bibinfo{person}{Xinyu Ma}, \bibinfo{person}{Pengjie Ren}, \bibinfo{person}{Dawei Yin}, {and} \bibinfo{person}{Zhaochun Ren}.} \bibinfo{year}{2023}\natexlab{}.
\newblock \showarticletitle{Is ChatGPT Good at Search? Investigating Large Language Models as Re-Ranking Agent}.
\newblock \bibinfo{journal}{\emph{ArXiv}}  \bibinfo{volume}{abs/2304.09542} (\bibinfo{year}{2023}).
\newblock


\bibitem[Tafjord et~al\mbox{.}(2019)]%
        {tafjord-etal-2019-quartz}
\bibfield{author}{\bibinfo{person}{Oyvind Tafjord}, \bibinfo{person}{Matt Gardner}, \bibinfo{person}{Kevin Lin}, {and} \bibinfo{person}{Peter Clark}.} \bibinfo{year}{2019}\natexlab{}.
\newblock \showarticletitle{{Q}ua{RT}z: An Open-Domain Dataset of Qualitative Relationship Questions}. In \bibinfo{booktitle}{\emph{Proceedings of the 2019 Conference on Empirical Methods in Natural Language Processing and the 9th International Joint Conference on Natural Language Processing (EMNLP-IJCNLP)}}, \bibfield{editor}{\bibinfo{person}{Kentaro Inui}, \bibinfo{person}{Jing Jiang}, \bibinfo{person}{Vincent Ng}, {and} \bibinfo{person}{Xiaojun Wan}} (Eds.). \bibinfo{publisher}{Association for Computational Linguistics}, \bibinfo{address}{Hong Kong, China}, \bibinfo{pages}{5941--5946}.
\newblock
\urldef\tempurl%
\url{https://doi.org/10.18653/v1/D19-1608}
\showDOI{\tempurl}


\bibitem[Talmor et~al\mbox{.}(2019)]%
        {commonsenseqa}
\bibfield{author}{\bibinfo{person}{Alon Talmor}, \bibinfo{person}{Jonathan Herzig}, \bibinfo{person}{Nicholas Lourie}, {and} \bibinfo{person}{Jonathan Berant}.} \bibinfo{year}{2019}\natexlab{}.
\newblock \showarticletitle{{C}ommonsense{QA}: A Question Answering Challenge Targeting Commonsense Knowledge}. In \bibinfo{booktitle}{\emph{Proceedings of the 2019 Conference of the North {A}merican Chapter of the Association for Computational Linguistics: Human Language Technologies, Volume 1 (Long and Short Papers)}}, \bibfield{editor}{\bibinfo{person}{Jill Burstein}, \bibinfo{person}{Christy Doran}, {and} \bibinfo{person}{Thamar Solorio}} (Eds.). \bibinfo{publisher}{Association for Computational Linguistics}, \bibinfo{address}{Minneapolis, Minnesota}, \bibinfo{pages}{4149--4158}.
\newblock
\urldef\tempurl%
\url{https://doi.org/10.18653/v1/N19-1421}
\showDOI{\tempurl}


\bibitem[Tamber et~al\mbox{.}(2023)]%
        {tamber2023scaling}
\bibfield{author}{\bibinfo{person}{Manveer~Singh Tamber}, \bibinfo{person}{Ronak Pradeep}, {and} \bibinfo{person}{Jimmy Lin}.} \bibinfo{year}{2023}\natexlab{}.
\newblock \showarticletitle{Scaling Down, LiTting Up: Efficient Zero-Shot Listwise Reranking with Seq2seq Encoder-Decoder Models}.
\newblock \bibinfo{journal}{\emph{arXiv preprint arXiv: 2312.16098}} (\bibinfo{year}{2023}).
\newblock


\bibitem[Tedeschi et~al\mbox{.}(2021)]%
        {wned}
\bibfield{author}{\bibinfo{person}{Simone Tedeschi}, \bibinfo{person}{Simone Conia}, \bibinfo{person}{Francesco Cecconi}, {and} \bibinfo{person}{Roberto Navigli}.} \bibinfo{year}{2021}\natexlab{}.
\newblock \showarticletitle{{N}amed {E}ntity {R}ecognition for {E}ntity {L}inking: {W}hat Works and What{'}s Next}. In \bibinfo{booktitle}{\emph{Findings of the Association for Computational Linguistics: EMNLP 2021}}. \bibinfo{publisher}{Association for Computational Linguistics}, \bibinfo{address}{Punta Cana, Dominican Republic}, \bibinfo{pages}{2584--2596}.
\newblock
\urldef\tempurl%
\url{https://aclanthology.org/2021.findings-emnlp.220}
\showURL{%
\tempurl}


\bibitem[Thorne et~al\mbox{.}(2018)]%
        {fever}
\bibfield{author}{\bibinfo{person}{James Thorne}, \bibinfo{person}{Andreas Vlachos}, \bibinfo{person}{Christos Christodoulopoulos}, {and} \bibinfo{person}{Arpit Mittal}.} \bibinfo{year}{2018}\natexlab{}.
\newblock \showarticletitle{{FEVER}: a Large-scale Dataset for Fact Extraction and {VERification}}. In \bibinfo{booktitle}{\emph{NAACL-HLT}}.
\newblock


\bibitem[Touvron et~al\mbox{.}(2023)]%
        {touvron2023llama}
\bibfield{author}{\bibinfo{person}{Hugo Touvron}, \bibinfo{person}{Louis Martin}, \bibinfo{person}{Kevin Stone}, \bibinfo{person}{Peter Albert}, \bibinfo{person}{Amjad Almahairi}, \bibinfo{person}{Yasmine Babaei}, \bibinfo{person}{Nikolay Bashlykov}, \bibinfo{person}{Soumya Batra}, \bibinfo{person}{Prajjwal Bhargava}, \bibinfo{person}{Shruti Bhosale}, {et~al\mbox{.}}} \bibinfo{year}{2023}\natexlab{}.
\newblock \showarticletitle{Llama 2: Open foundation and fine-tuned chat models}.
\newblock \bibinfo{journal}{\emph{arXiv preprint arXiv:2307.09288}} (\bibinfo{year}{2023}).
\newblock


\bibitem[Trivedi et~al\mbox{.}(2022)]%
        {musique}
\bibfield{author}{\bibinfo{person}{Harsh Trivedi}, \bibinfo{person}{Niranjan Balasubramanian}, \bibinfo{person}{Tushar Khot}, {and} \bibinfo{person}{Ashish Sabharwal}.} \bibinfo{year}{2022}\natexlab{}.
\newblock \showarticletitle{{M}u{S}i{Q}ue: Multihop Questions via Single-hop Question Composition}.
\newblock \bibinfo{journal}{\emph{Transactions of the Association for Computational Linguistics}} (\bibinfo{year}{2022}).
\newblock


\bibitem[Wang et~al\mbox{.}(2024)]%
        {wang2024utilizing}
\bibfield{author}{\bibinfo{person}{Jiajia Wang}, \bibinfo{person}{Jimmy~Xiangji Huang}, \bibinfo{person}{Xinhui Tu}, \bibinfo{person}{Junmei Wang}, \bibinfo{person}{Angela~Jennifer Huang}, \bibinfo{person}{Md~Tahmid~Rahman Laskar}, {and} \bibinfo{person}{Amran Bhuiyan}.} \bibinfo{year}{2024}\natexlab{}.
\newblock \showarticletitle{Utilizing BERT for Information Retrieval: Survey, Applications, Resources, and Challenges}.
\newblock \bibinfo{journal}{\emph{Comput. Surveys}} \bibinfo{volume}{56}, \bibinfo{number}{7} (\bibinfo{year}{2024}), \bibinfo{pages}{1--33}.
\newblock


\bibitem[Wang et~al\mbox{.}(2022a)]%
        {wang2022archivalqa}
\bibfield{author}{\bibinfo{person}{Jiexin Wang}, \bibinfo{person}{Adam Jatowt}, {and} \bibinfo{person}{Masatoshi Yoshikawa}.} \bibinfo{year}{2022}\natexlab{a}.
\newblock \showarticletitle{Archivalqa: A large-scale benchmark dataset for open-domain question answering over historical news collections}. In \bibinfo{booktitle}{\emph{Proceedings of the 45th International ACM SIGIR Conference on Research and Development in Information Retrieval}}. \bibinfo{pages}{3025--3035}.
\newblock


\bibitem[Wang et~al\mbox{.}(2020)]%
        {wang2020minilmv2}
\bibfield{author}{\bibinfo{person}{Wenhui Wang}, \bibinfo{person}{Hangbo Bao}, \bibinfo{person}{Shaohan Huang}, \bibinfo{person}{Li Dong}, {and} \bibinfo{person}{Furu Wei}.} \bibinfo{year}{2020}\natexlab{}.
\newblock \showarticletitle{Minilmv2: Multi-head self-attention relation distillation for compressing pretrained transformers}.
\newblock \bibinfo{journal}{\emph{arXiv preprint arXiv:2012.15828}} (\bibinfo{year}{2020}).
\newblock


\bibitem[Wang et~al\mbox{.}(2022b)]%
        {wang2022self}
\bibfield{author}{\bibinfo{person}{Xuezhi Wang}, \bibinfo{person}{Jason Wei}, \bibinfo{person}{Dale Schuurmans}, \bibinfo{person}{Quoc Le}, \bibinfo{person}{Ed Chi}, \bibinfo{person}{Sharan Narang}, \bibinfo{person}{Aakanksha Chowdhery}, {and} \bibinfo{person}{Denny Zhou}.} \bibinfo{year}{2022}\natexlab{b}.
\newblock \showarticletitle{Self-consistency improves chain of thought reasoning in language models}.
\newblock \bibinfo{journal}{\emph{arXiv preprint arXiv:2203.11171}} (\bibinfo{year}{2022}).
\newblock


\bibitem[Xiao et~al\mbox{.}(2023)]%
        {bge_embedding}
\bibfield{author}{\bibinfo{person}{Shitao Xiao}, \bibinfo{person}{Zheng Liu}, \bibinfo{person}{Peitian Zhang}, {and} \bibinfo{person}{Niklas Muennighoff}.} \bibinfo{year}{2023}\natexlab{}.
\newblock \bibinfo{title}{C-Pack: Packaged Resources To Advance General Chinese Embedding}.
\newblock
\newblock
\showeprint[arxiv]{2309.07597}~[cs.CL]


\bibitem[Xiong et~al\mbox{.}(2024)]%
        {xiong2024search}
\bibfield{author}{\bibinfo{person}{Haoyi Xiong}, \bibinfo{person}{Jiang Bian}, \bibinfo{person}{Yuchen Li}, \bibinfo{person}{Xuhong Li}, \bibinfo{person}{Mengnan Du}, \bibinfo{person}{Shuaiqiang Wang}, \bibinfo{person}{Dawei Yin}, {and} \bibinfo{person}{Sumi Helal}.} \bibinfo{year}{2024}\natexlab{}.
\newblock \showarticletitle{When search engine services meet large language models: visions and challenges}.
\newblock \bibinfo{journal}{\emph{IEEE Transactions on Services Computing}} (\bibinfo{year}{2024}).
\newblock


\bibitem[Xiong et~al\mbox{.}(2021)]%
        {ance}
\bibfield{author}{\bibinfo{person}{Lee Xiong}, \bibinfo{person}{Chenyan Xiong}, \bibinfo{person}{Ye Li}, \bibinfo{person}{Kwok-Fung Tang}, \bibinfo{person}{Jialin Liu}, \bibinfo{person}{Paul~N. Bennett}, \bibinfo{person}{Junaid Ahmed}, {and} \bibinfo{person}{Arnold Overwijk}.} \bibinfo{year}{2021}\natexlab{}.
\newblock \showarticletitle{Approximate Nearest Neighbor Negative Contrastive Learning for Dense Text Retrieval}. In \bibinfo{booktitle}{\emph{Proceedings of the 9th International Conference on Learning Representations (ICLR 2021)}}.
\newblock


\bibitem[Yamada et~al\mbox{.}(2021)]%
        {yamada2021bpr}
\bibfield{author}{\bibinfo{person}{Ikuya Yamada}, \bibinfo{person}{Akari Asai}, {and} \bibinfo{person}{Hannaneh Hajishirzi}.} \bibinfo{year}{2021}\natexlab{}.
\newblock \showarticletitle{Efficient Passage Retrieval with Hashing for Open-domain Question Answering}. In \bibinfo{booktitle}{\emph{ACL}}.
\newblock


\bibitem[Yang et~al\mbox{.}(2019)]%
        {yang2019hybrid}
\bibfield{author}{\bibinfo{person}{Liu Yang}, \bibinfo{person}{Junjie Hu}, \bibinfo{person}{Minghui Qiu}, \bibinfo{person}{Chen Qu}, \bibinfo{person}{Jianfeng Gao}, \bibinfo{person}{W~Bruce Croft}, \bibinfo{person}{Xiaodong Liu}, \bibinfo{person}{Yelong Shen}, {and} \bibinfo{person}{Jingjing Liu}.} \bibinfo{year}{2019}\natexlab{}.
\newblock \showarticletitle{A hybrid retrieval-generation neural conversation model}. In \bibinfo{booktitle}{\emph{Proceedings of the 28th ACM international conference on information and knowledge management}}. \bibinfo{pages}{1341--1350}.
\newblock


\bibitem[Yang et~al\mbox{.}(2015)]%
        {wikiqa}
\bibfield{author}{\bibinfo{person}{Yi Yang}, \bibinfo{person}{Wen-tau Yih}, {and} \bibinfo{person}{Christopher Meek}.} \bibinfo{year}{2015}\natexlab{}.
\newblock \showarticletitle{{W}iki{QA}: A Challenge Dataset for Open-Domain Question Answering}. In \bibinfo{booktitle}{\emph{Proceedings of the 2015 Conference on Empirical Methods in Natural Language Processing}}, \bibfield{editor}{\bibinfo{person}{Llu{\'\i}s M{\`a}rquez}, \bibinfo{person}{Chris Callison-Burch}, {and} \bibinfo{person}{Jian Su}} (Eds.). \bibinfo{publisher}{Association for Computational Linguistics}, \bibinfo{address}{Lisbon, Portugal}, \bibinfo{pages}{2013--2018}.
\newblock
\urldef\tempurl%
\url{https://doi.org/10.18653/v1/D15-1237}
\showDOI{\tempurl}


\bibitem[Yang et~al\mbox{.}(2018)]%
        {hotpotqa}
\bibfield{author}{\bibinfo{person}{Zhilin Yang}, \bibinfo{person}{Peng Qi}, \bibinfo{person}{Saizheng Zhang}, \bibinfo{person}{Yoshua Bengio}, \bibinfo{person}{William Cohen}, \bibinfo{person}{Ruslan Salakhutdinov}, {and} \bibinfo{person}{Christopher~D. Manning}.} \bibinfo{year}{2018}\natexlab{}.
\newblock \showarticletitle{{H}otpot{QA}: A Dataset for Diverse, Explainable Multi-hop Question Answering}. In \bibinfo{booktitle}{\emph{Proceedings of the 2018 Conference on Empirical Methods in Natural Language Processing}}, \bibfield{editor}{\bibinfo{person}{Ellen Riloff}, \bibinfo{person}{David Chiang}, \bibinfo{person}{Julia Hockenmaier}, {and} \bibinfo{person}{Jun{'}ichi Tsujii}} (Eds.). \bibinfo{publisher}{Association for Computational Linguistics}, \bibinfo{address}{Brussels, Belgium}, \bibinfo{pages}{2369--2380}.
\newblock
\urldef\tempurl%
\url{https://doi.org/10.18653/v1/D18-1259}
\showDOI{\tempurl}


\bibitem[Yates et~al\mbox{.}(2021)]%
        {dpr}
\bibfield{author}{\bibinfo{person}{Andrew Yates}, \bibinfo{person}{Rodrigo Nogueira}, {and} \bibinfo{person}{Jimmy Lin}.} \bibinfo{year}{2021}\natexlab{}.
\newblock \showarticletitle{Pretrained transformers for text ranking: BERT and beyond}. In \bibinfo{booktitle}{\emph{Proceedings of the 14th ACM International Conference on web search and data mining}}. \bibinfo{pages}{1154--1156}.
\newblock


\bibitem[Yoon et~al\mbox{.}(2024)]%
        {yoon2024listt5listwisererankingfusionindecoder}
\bibfield{author}{\bibinfo{person}{Soyoung Yoon}, \bibinfo{person}{Eunbi Choi}, \bibinfo{person}{Jiyeon Kim}, \bibinfo{person}{Hyeongu Yun}, \bibinfo{person}{Yireun Kim}, {and} \bibinfo{person}{Seung won Hwang}.} \bibinfo{year}{2024}\natexlab{}.
\newblock \bibinfo{title}{ListT5: Listwise Reranking with Fusion-in-Decoder Improves Zero-shot Retrieval}.
\newblock
\newblock
\showeprint[arxiv]{2402.15838}~[cs.IR]
\urldef\tempurl%
\url{https://arxiv.org/abs/2402.15838}
\showURL{%
\tempurl}


\bibitem[Yu et~al\mbox{.}(2024)]%
        {yu2024localrqa}
\bibfield{author}{\bibinfo{person}{Xiao Yu}, \bibinfo{person}{Yunan Lu}, {and} \bibinfo{person}{Zhou Yu}.} \bibinfo{year}{2024}\natexlab{}.
\newblock \showarticletitle{LocalRQA: From Generating Data to Locally Training, Testing, and Deploying Retrieval-Augmented QA Systems}.
\newblock \bibinfo{journal}{\emph{arXiv preprint arXiv:2403.00982}} (\bibinfo{year}{2024}).
\newblock


\bibitem[Zellers et~al\mbox{.}(2019)]%
        {hellaswag}
\bibfield{author}{\bibinfo{person}{Rowan Zellers}, \bibinfo{person}{Ari Holtzman}, \bibinfo{person}{Yonatan Bisk}, \bibinfo{person}{Ali Farhadi}, {and} \bibinfo{person}{Yejin Choi}.} \bibinfo{year}{2019}\natexlab{}.
\newblock \showarticletitle{HellaSwag: Can a Machine Really Finish Your Sentence?}. In \bibinfo{booktitle}{\emph{Proceedings of the 57th Annual Meeting of the Association for Computational Linguistics}}.
\newblock


\bibitem[Zeng et~al\mbox{.}(2022)]%
        {cldrd}
\bibfield{author}{\bibinfo{person}{Hansi Zeng}, \bibinfo{person}{Hamed Zamani}, {and} \bibinfo{person}{Vishwa Vinay}.} \bibinfo{year}{2022}\natexlab{}.
\newblock \showarticletitle{Curriculum Learning for Dense Retrieval Distillation}. In \bibinfo{booktitle}{\emph{Proc. SIGIR}}. \bibinfo{pages}{1979–1983}.
\newblock


\bibitem[Zhu et~al\mbox{.}(2021)]%
        {zhu2021leveraging}
\bibfield{author}{\bibinfo{person}{Nengjun Zhu}, \bibinfo{person}{Jian Cao}, \bibinfo{person}{Xinjiang Lu}, {and} \bibinfo{person}{Qi Gu}.} \bibinfo{year}{2021}\natexlab{}.
\newblock \showarticletitle{Leveraging pointwise prediction with learning to rank for top-N recommendation}.
\newblock \bibinfo{journal}{\emph{World Wide Web}}  \bibinfo{volume}{24} (\bibinfo{year}{2021}), \bibinfo{pages}{375--396}.
\newblock


\bibitem[Zhuang et~al\mbox{.}(2023)]%
        {zhuang2023rankt5}
\bibfield{author}{\bibinfo{person}{Honglei Zhuang}, \bibinfo{person}{Zhen Qin}, \bibinfo{person}{Rolf Jagerman}, \bibinfo{person}{Kai Hui}, \bibinfo{person}{Ji Ma}, \bibinfo{person}{Jing Lu}, \bibinfo{person}{Jianmo Ni}, \bibinfo{person}{Xuanhui Wang}, {and} \bibinfo{person}{Michael Bendersky}.} \bibinfo{year}{2023}\natexlab{}.
\newblock \showarticletitle{Rankt5: Fine-tuning t5 for text ranking with ranking losses}. In \bibinfo{booktitle}{\emph{Proceedings of the 46th International ACM SIGIR Conference on Research and Development in Information Retrieval}}. \bibinfo{pages}{2308--2313}.
\newblock


\end{thebibliography}

\end{document}